\documentclass[aps,prx,10pt,superscriptaddress,floatfix]{revtex4-2}
\usepackage[utf8]{inputenc}
\usepackage{ulem}
\usepackage[T1]{fontenc}
\usepackage{lmodern}
\usepackage{iftex}
\usepackage{braket}
\usepackage[version=3]{mhchem}
\usepackage{multirow}
\usepackage{longtable}
\usepackage{ascmac}
\ifPDFTeX
\usepackage{color}
\usepackage{xcolor}
\usepackage{graphicx}
\else
\usepackage[dvipdfmx]{color}
\usepackage[dvipdfmx]{xcolor}
\usepackage[dvipdfmx]{graphicx}
\fi
\usepackage{ulem}
\usepackage{booktabs} 

\usepackage{natbib}
\usepackage{mathtools}
\usepackage{amsfonts,amsmath,amssymb,amsthm}
\usepackage{silence}
\usepackage{dblfloatfix}
\ifPDFTeX
\usepackage[hypertexnames=false]{hyperref}
\else
\usepackage[dvipdfmx,hypertexnames=false]{hyperref}
\fi
\theoremstyle{plain}

\newtheorem*{thm*}{Theorem}

\usepackage{bm}
\usepackage{bbm}

\newcommand{\bx}{\ensuremath{\boldsymbol{x}}}

\newcommand{\Ctot}{\ensuremath{C_{\rm tot}}}

\makeatletter
\newcommand{\figcaption}[1]{\def\@captype{figure}\caption{#1}}
\newcommand{\tblcaption}[1]{\def\@captype{table}\caption{#1}}
\makeatother

\begin{document}
\title{
Quantum reservoir computing with repeated measurements
\texorpdfstring{\\}{ }
on superconducting devices
}

\author{Toshiki Yasuda}
\affiliation{Department of Applied Physics and Physico-Informatics, Keio University, 3-14-1 Hiyoshi, Kohoku, Yokohama, 223-8522, Japan}
\author{Yudai Suzuki}
\affiliation{Institute of Physics, Ecole Polytechnique Fédérale de Lausanne (EPFL), Lausanne, Switzerland}
\affiliation{Centre for Quantum Science and Engineering, Ecole Polytechnique Fédérale de Lausanne (EPFL), Lausanne, Switzerland}
\affiliation{Quantum Computing Center, Keio University, Hiyoshi 3-14-1, Kohoku, Yokohama 223-8522, Japan}
\author{Tomoyuki Kubota}
\affiliation{Graduate School of Information Science and Technology, The University of Tokyo,\\
7-3-1 Hongo, Bunkyo-ku, Tokyo, 113-8556, Japan}
\author{Kohei Nakajima}
\affiliation{Graduate School of Information Science and Technology, The University of Tokyo,\\
7-3-1 Hongo, Bunkyo-ku, Tokyo, 113-8556, Japan}
\author{Qi Gao}
\affiliation{Mitsubishi Chemical Corporation, Science \& Innovation Center, 1000, Kamoshida-cho, Aoba-ku, Yokohama 227-8502, Japan}
\affiliation{Quantum Computing Center, Keio University, Hiyoshi 3-14-1, Kohoku, Yokohama 223-8522, Japan}
\author{Wenlong Zhang}
\affiliation{School of Manufacturing Systems and Networks, Arizona State University, Mesa, AZ 85212, USA}
\author{Satoshi Shimono}
\affiliation{The Global KAITEKI Center, Arizona State University, Tempe, AZ 85281, USA}
\author{Atsushi Matsuo}
\affiliation{IBM Research, Tokyo 103-8510, Japan}
\author{Janani Ananthanarayanan}
\affiliation{IBM Research, Bengaluru, India}
\author{Hendra I. Nurdin}
\affiliation{School of Electrical Engineering and Telecommunications, The University of New South Wales, \\ Sydney, New South Wales 2052, Australia}
\author{Naoki Yamamoto}
\affiliation{Department of Applied Physics and Physico-Informatics, Keio University, 3-14-1 Hiyoshi, Kohoku, Yokohama, 223-8522, Japan}
\affiliation{Quantum Computing Center, Keio University, Hiyoshi 3-14-1, Kohoku, Yokohama 223-8522, Japan}

\begin{abstract}
Reservoir computing is a machine learning framework that uses artificial or
physical dissipative dynamics to predict time-series data using nonlinearity
and memory properties of dynamical systems.
Quantum systems are considered promising physical reservoirs, and several quantum reservoir computing (QRC) schemes that exploit intrinsic quantum mechanical properties as computational resources have been proposed.
This paper focuses on a particular QRC scheme that was theoretically proposed to use the repeated quantum non-demolition measurement to generate a time series, mainly for the purpose of reducing the total execution time compared to the conventional method without repeated measurement.
We experimentally implemented this scheme on IBM superconducting devices and demonstrated that shorter execution time and higher accuracy can be achieved than with a specific conventional method.
Then, we study the temporal information processing capacity to quantify the computational capability of the proposed QRC; in particular, we use this
quantity to identify the measurement strength that finds the best balance between the amount of available information and the strength of dissipation.
An experimental demonstration with a soft robot is also provided, where repeated measurement over 1,000 timesteps was effectively implemented.
Finally, we implemented the repeated-measurement QRC scheme using 120 qubits, to test the potential computational power of a large-scale QRC.
\end{abstract}

\maketitle

\section{Introduction}
\label{sec:intro}

\begin{figure*}[tp]
    \centering
    \begin{center}
     \includegraphics[width=\textwidth]{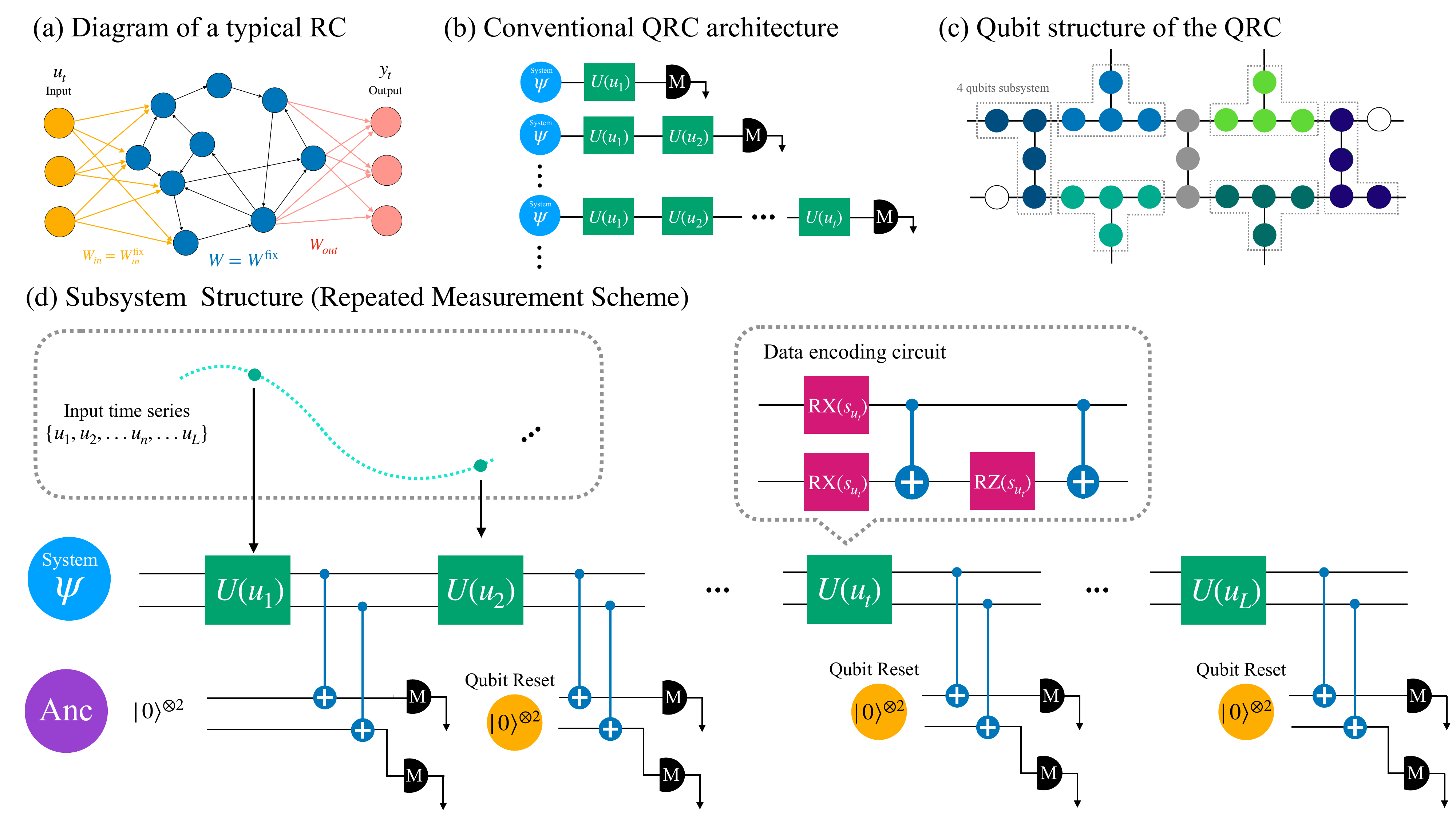}
    \end{center}
        \caption{
        (a) Diagram of a typical reservoir computing system. The input passes through an intermediate artificial or physical layer, and a linear readout is applied to the reservoir output. The weights of the intermediate layer are fixed and are not trained.
        (b) Conventional QRC model without repeated QND measurements. To obtain the output signal at time \(t\), a circuit evolved up to time \(t\) is prepared, and measurements are performed at the end of the circuit to estimate the expectation values.
        (c) Qubit arrangement on ibm\_kingston. The four qubits indicated by each gray dotted box constitute one QRC component, composed of a two-qubit system and a two-qubit ancilla register.
        (d) Detailed structure of the four-qubit component of our QRC model. The two-qubit system encodes the input time series, and the two-qubit ancilla register is measured to produce the outputs. Each system unitary \(U(u_t)\), indicated by the gray dashed line, consists of RX, RZ, and CNOT gates and encodes the input time series into rotation angles. The system and ancilla are coupled by two CNOT gates. The ancilla state is reset to \(|0\rangle^{\otimes 2}\) after each measurement. 
        In the actual device there is decoherence so that the operation actually implemented at time $t$ is not $U(u_t)$ but a version of it undergoing decoherence due to noise in the quantum system; this is not explicitly depicted in the sub-figures (b)-(d) above.}
    \label{fig:architecture}
\end{figure*}

Quantum computing is a next-generation computing technology that will bring
about possible game-changing computing capabilities in various fields.
Since currently available quantum devices are noisy and intermediate-scale, quantum
computing algorithms that can run even on such devices have been actively investigated.
In particular, there are many proposals for machine learning applications, many of which are based on variational quantum algorithms (VQAs)~\cite{VQA}, where parametrized
quantum circuits are trained analogously to neural networks.
However, parametrized quantum circuits are generally difficult to train \cite{BP, Cerezo2021, Wang2021, Ragone2024},
while several techniques to circumvent this issue have also been developed;
e.g., see Refs.~\cite{Grant2019,Skolik2021,Rudolph2023}.

With such motivation, quantum reservoir computing (QRC)
has emerged as a promising route toward quantum machine learning~\cite{QR, BoostingQRC2019, Fujii2021}.
Reservoir computing (RC) is a general machine learning framework that is often used for approximating time-series functions~\cite{RC, ESN, LSM, uni_rc, ESN_LSM}.
A major advantage of this method is its significantly lower training cost compared to (deep) neural-network-based machine-learning schemes.
This is because the learning process is performed through simple linear regression on the network outputs, while the internal dynamics of the network remains fixed. Namely, in contrast to the training of neural networks that generally requires optimization of all parameters in the input, hidden, and output layers, the RC scheme learns only the parameters in the output layer, while keeping the (input and) reservoir dynamics fixed during training; see Fig.~\ref{fig:architecture}(a).
Here, the fixed internal dynamics is referred to as the reservoir, which should have
sufficient computational capability to extract features for reproducing a target time series approximately.
Therefore, the choice of reservoir is critical for the success of RC.
Thus far, numerous studies have been conducted to identify what types of reservoirs are effective computational resources; for example, neural-network-based architectures such as echo state networks (ESNs) have been studied.
In particular, the RC paradigm with physical reservoirs, called physical RC, can potentially enhance the expressivity of the models with fewer computational resources by leveraging the high nonlinearity inherent in physical systems. Examples of such physical RC frameworks include soft matter \cite{softrobo, Bhovad2021, muscle2022}, spintronics \cite{spin, Furuta2018, Tsunegi2018, Akashi2020}, and electronic devices \cite{MERKEL2014, memori, Zhong2021}.

The QRC is a variant of physical RC models, where quantum systems are used as reservoirs.
Thanks to their intrinsic complexity, quantum systems are considered promising reservoirs \cite{QRC, qrc_rev, Sim1, Sim2, use_disspation_0}, as demonstrated in experiments on
superconducting devices \cite{ibm-qnd, transmon, naturalQRC, quantum_noise, ibm-reset} and nuclear magnetic resonance \cite{NMR}.
In addition, QRC has been exploited for some applications such as fluid analysis \cite{flow}, image recognition \cite{image}, and position estimation of mobile wireless device users \cite{mobile}. Furthermore, the power of QRC has been studied in light of quantum resources such as entanglement, indicating the possibility that quantum reservoir schemes might be intrinsically superior to classical reservoir schemes~\cite{entangle}.

Among previous studies that use quantum computers as reservoirs, Refs.~\cite{ibm-qnd, transmon, naturalQRC, quantum_noise, ibm-reset} implemented quantum reservoirs on actual IBM superconducting quantum devices, with a specific focus on the use of intrinsic noise on quantum hardware. 
Note that, in the conventional QRC scheme (i.e., QRC without repeated measurement), the system is re-prepared and run from the beginning for each timestep to produce the output of the entire time series, which is obtained by averaging the measurement results on all qubits at each time; see Fig.~\ref{fig:architecture}(b). This is because performing strong measurements, such as projective measurements, to extract output from the system causes the loss of input information and coherence. 
The process of re-preparing the system from the initial state not only requires storing many inputs and quantum circuits in memory but also clearly requires a long execution time.

An approach to mitigating this challenge is the quantum non-demolition (QND)
measurement, which was theoretically introduced into the QRC scheme in
Ref.~\cite[\S V and Appendix C]{ibm-qnd}. Ref.~\cite{ibm-qnd} first proposed
the use of repeated QND measurements in the context of QRC, but did not provide
a quantitative experimental analysis or a hardware implementation since mid-circuit measurements were not yet available at the time on the IBM cloud computer used in \cite{ibm-qnd}. 

The contribution of this paper is to show an experimental demonstration of the QRC scheme based on the natural QRC scheme from \cite{naturalQRC} but incorporating the repeated QND measurement proposed in Ref.~\cite{ibm-qnd} on IBM superconducting quantum devices and investigate its computational properties. Note that \cite{ibm-qnd} considers a different dynamical QRC model from \cite{naturalQRC} as emphasized further below. 
In this work we do not use the dynamical quantum model proposed in \cite{ibm-qnd}, which is dissipative rather than unitary and with a built-in probabilistic reset feature that gives the model the so-called convergence or echo state property central to reservoir computing. Instead we implement the simpler natural QRC model from \cite{naturalQRC}, in which a unitary evolution is implemented but dissipativity is provided by the decoherence due to the natural noise in a device.

Fig.~\ref{fig:architecture}(d) depicts the QRC structure we implemented, for the case of a two-qubit system and a two-qubit ancilla.
This scheme repeatedly performs a QND measurement on additional ancillas that are prepared and coupled to the system qubits at each time, and obtains one stochastic time series through a single run of the entire system. 
This procedure is repeated to obtain a set of stochastic time series; they are finally averaged to produce one deterministic time series with finite sampling errors that depend on the number of samples used for averaging. 
As a result, only a single quantum circuit and input sequence are required, leading to a shortening of the overall execution time of the QRC.

Note that Ref.~\cite{QRC-weak} theoretically proposed a similar yet general repeated-measurement protocol in the sense that it does not assume the use of a specific quantum system or a specific measurement implementation. 
In Ref.~\cite{QRC-weak}, three protocols are considered: Restart Protocol (RSP), Online Protocol (OLP), and Rewinding Protocol (RWP). These protocols were previously proposed in \cite{chen2019towards} (RSP), \cite[Section V]{ibm-qnd} (RWP) and \cite[Section V and Appendix C]{ibm-qnd} (OLP). The method proposed in Ref.~\cite{naturalQRC} corresponds to RSP, while our method corresponds to OLP. 
For these protocols, roughly speaking, the time complexities are represented as $t_{\rm RSP} \in \mathcal{O}(N_SL^2)$, $t_{\rm RWP} \in \mathcal{O}(N_SLk)$ and $t_{\rm OLP} \in \mathcal{O}(N_SL)$, respectively~\cite[arXiv version, Appendix G]{QRC-weak}.
Here, \(N_S\) represents the number of shots, \(L\) is the total length of the time series, and \(k\) is the size of the time window.
 From this standpoint, Ref.~\cite{QRC-weak} theoretically analyzed measurement-based QRC protocols and numerically investigated their resource efficiency and performance.

Another remark is that, recently in \cite{Coherence}, it was shown that in the ideal situation where the circuit shown in Fig.~\ref{fig:architecture}(d) is implemented perfectly (i.e., in the absence of any decoherence in particular meaning), it would undergo thermalization because the underlying CPTP map on the system qubits induced by the system--ancilla interaction at a single timestep would be unital. 
A consequence of this is that the evolution of the quantum state is independent of the input sequence. That is, the quantum state does not encode any information about the input sequence. However, our circuit is not ideal and there is decoherence. As shown in Section \ref{sec:results}, our experimental results indicate that our circuit does not undergo thermalization and is able to perform processing of input sequences. We believe that this can be attributed to the particular decoherence in the circuit results in a CPTP map that is not unital and helps to avoid the predicted thermalization or at least is able to slow it down so that our QRC circuit is still able to map input sequences to quantum states.

We also emphasize that a potential advantage of our scheme is that quantum computers with a large number of qubits can be used for the real-time execution of QRC.
Many of the leading architectures for quantum computing and modern quantum algorithms anticipate the use of millions of qubits, for instance with the use of quantum error correction based on the surface code~\cite{surfac-code-QEC}.
As a concrete example, suppose that 10,000 identical copies of an $N$-qubit system can be realized. If 200 independent circuits running with repeated QND measurements can be executed on each copy within $10^{-5}$ seconds, then averaging over the measurements would in principle allow for an $N$-qubit QRC with a high-accuracy estimation of qubit expectation values comparable to an 8-bit AD converter running at a sampling rate of 100 kHz. With more copies available, even higher-accuracy estimation could be aimed for.
However, continuous running of the QRC in this setting requires the convergence property as in the QRC architecture in Ref.~\cite{ibm-qnd} so that it asymptotically forgets its initial state.
This allows QND measurements to be run only over a moving window of finite length as discussed in Ref.~\cite[\S V and Appendix D]{ibm-qnd}.

The results obtained in this paper are as follows.
Firstly, we study the task of emulating the Nonlinear Auto-Regressive Moving Average dynamics (NARMA), to compare the performance of our repeated-measurement QRC with the conventional QRC scheme, particularly the natural-noise scheme in Ref.~\cite{naturalQRC}, which does not use repeated measurement.
We remark that NARMA is known as a benchmark task for temporal information processing. Next, to examine the practical applicability of our scheme, we study the task of emulating the time-series data obtained from the experiment for soft robotics.
Notably, in this experiment, we conducted repeated QND measurements over 1,000 timesteps, which indicates that our scheme would be available for real applications.

Moreover, we calculate the temporal information processing capacity (TIPC)
\cite{IPC_calc, quantum_noise} to analyze the computational capability of our
repeated-measurement QRC.
TIPC is known as a quantity that can comprehensively reveal the computational capability of general reservoir systems by quantifying the amount of time-invariant and time-variant terms included in the state.
Note that TIPC does not depend on a specific signal processing task but is intrinsic to the reservoir.
RC requires reservoir dynamics to hold an echo state property, under which its state is a function of only input history (called time-invariant).
The time-invariant state is required to form an output reproducible against an identical input sequence; on the other hand, sometimes, the quantum reservoir state can include not only time-invariant terms but also time-variant ones (i.e., a function of both time and input history) \cite{quantum_noise}, and all of the terms cannot be utilized for computation.
For the linear regression
problem, time-invariant memories contribute to the performance of the task but time-variant ones cannot
improve it; this is because, while the target output is a time-invariant function, the linear readout just
multiplies coefficients by the reservoir state and thus cannot remove the time dependence of the state. For this reason, while conventional RC focuses on only the time-invariant terms, generalized reservoir computing \cite{genRC} can potentially exploit not only time-invariant but also time-variant terms to improve the performance with a properly adjusted readout.
Thus, following the previous work~\cite{quantum_noise}, we used the quantity for the analysis.
Also, we calculated the TIPC as a function of the interaction strength between the system and the ancilla, confirming that the time-invariant capacity has its highest
value at an intermediate interaction strength. This is the best trade-off point
between the amount of available information and the strength of dissipation; the existence of this point is consistent with the result of Ref.~\cite{QRC-weak}.
This is a clear advantage of our QRC scheme in that it
enables tuning such an important trade-off parameter, which is not available in
the natural-noise scheme.

Lastly, we conducted some experiments on QRC using 120 qubits (more precisely, 90 system qubits and 30 ancilla qubits) and discussed the potential capability of QRC with a large quantum reservoir system that may be impossible to simulate on classical computers.
More specifically, we conducted the NARMA2 experiment on four types of 120-qubit devices, along with calculating the TIPC.
Also, we performed the NARMA2 experiment where the output results of multiple 120-qubit devices are combined in both spatial and time directions, to artificially enlarge the system size.
In all results, the 120-qubit repeated-measurement QRC did not outperform the smaller-size one, although the results provide some insights and suggestions toward realizing large-size QRC in the future.

\section{Method}
\subsection{QRC with repeated measurement}

Our QRC model consists of an \(n\)-qubit system and an \(n\)-qubit ancilla register.
Its one-step evolution is described by an input-dependent quantum channel together with a projective measurement on the ancilla register, repeated at each timestep, as follows:
\begin{equation}
\label{system dynamics}
     \rho_{t}^{(\bm{m}_{t})}
        = \frac{1}{p(\bm{m}_{t})}
            \mathrm{Tr}_{a}\Bigl[M_{\bm{m}_t}\Phi_{u_t}(\rho_{t-1}^{(\bm{m}_{t-1})}\otimes \sigma_a) {M^{\dagger}_{\bm{m}_t}}\Bigl],
\end{equation}
where $\sigma_a=(|0\rangle_a\langle0|)^{\otimes n}$ and
\begin{equation}
\label{proj meas}
     M_{\bm{m}_t}
       = I_{s} \otimes \bigotimes_{i=1}^n|m_{i,t}\rangle_a \langle m_{i,t}|.
\end{equation}
Here, \(\rho_t^{(\bm{m}_t)}\) is the system density matrix at timestep \(t\), conditioned on the measurement outcome \(\bm{m}_t\).
The one-step evolution is described by the input-dependent quantum channel \(\Phi_{u_t}\), which includes both the intended unitary dynamics and noise effects associated with the NISQ device.
More explicitly, we write
\[
\Phi_{u_t} = \mathcal{T} \circ \mathcal{U}_{u_t},
\qquad
\mathcal{U}_{u_t}(\,\cdot\,) = \hat{U}(u_t)(\,\cdot\,)\hat{U}^\dagger(u_t),
\]
where \(\mathcal{T}\) denotes a completely positive trace-preserving (CPTP) map representing decoherence and other noise processes, such as gate-dependent noise and crosstalk.
The repeated measurement is implemented as a computational-basis projective measurement on the ancilla register, as given in Eq.~\eqref{proj meas}, where \(m_{i,t}\in\{0,1\}\) represents the binary measurement outcome on the \(i\)-th ancilla qubit, and the bit string \(\bm{m}_t = m_{1,t}\cdots m_{n,t}\) denotes the measurement outcome at timestep \(t\).
The corresponding probability of obtaining \(\bm{m}_t\) is
\[
p(\bm{m}_t)=\mathrm{Tr}\!\left[
M_{\bm{m}_t}\,
\Phi_{u_t}\!\left(\rho_{t-1}^{(\bm{m}_{t-1})}\otimes\sigma_a\right)
M_{\bm{m}_t}^\dagger
\right].
\]
Thus, the conditional state of the QRC system evolves stochastically, depending on the measurement outcome of the ancilla register.
Fig.~\ref{fig:architecture}(d) illustrates the circuit realization of our model on an IBM superconducting quantum processor.

The measurement results are used to construct the output of the quantum reservoir system.
Specifically, the output vector at timestep $t$ is composed of the expectation values of the Pauli $Z$ matrices on the ancilla qubits:
\begin{equation}
\boldsymbol{h}_{t}
   = [\langle Z_{1,a} \rangle, \langle Z_{2,a} \rangle, \ldots , \langle Z_{n,a} \rangle]^T,
\end{equation}
where $Z_{i,a}$ is the Pauli $Z$ matrix of the $i$-th qubit of the ancilla system,
i.e., $Z_{i,a} = I\otimes \cdots \otimes Z \otimes \cdots \otimes I$. 
We approximately obtain each element of the output  $\boldsymbol{h}_t$ by repeating the experiment $N_s$ times and averaging the $N_s$ bit strings $B_{t}=\{\bm{m}^{(1)}_{t},\bm{m}^{(2)}_{t},\ldots, \bm{m}^{(N_{s})}_{t}\}$ as follows:
\begin{equation}
\label{linear dynamics}
\begin{aligned}
\langle Z_{i,a} \rangle
&=
\sum_{\bm{m'}_{t-1}\in\{0,1\}^n}
p(\bm{m'}_{t-1})\, \times
\mathrm{Tr}\!\left[
Z_{i,a}\,
\Phi_{u_t}\!\left(
\rho_{t-1}^{(\bm{m'}_{t-1})}\otimes \sigma_a
\right)
\right] \\
&= \mathrm{Tr}\!\left[
Z_{i,a}\,
\Phi_{u_t}\!\left(
\sum_{\bm{m'}_{t-1}\in\{0,1\}^n}
p(\bm{m'}_{t-1})\,\rho_{t-1}^{(\bm{m'}_{t-1})}
\otimes \sigma_a
\right)
\right] \\
&\approx \frac{1}{N_s}\sum_{\bm{m}_k\in B_t}\mathbbm{1}_{B_{i,0}}[\bm{m}_k]
- \frac{1}{N_s}\sum_{\bm{m}_k\in B_t}\mathbbm{1}_{B_{i,1}}[\bm{m}_k],
\end{aligned}
\end{equation}
where $\mathbbm{1}_A[\cdot]$ is the indicator function and
$B_{i,l}=\{\bm{m} \mid \bm{m}\in B_t,\; m_i=l\}$.
In our demonstration shown later, the number of repetitions used to approximate the output values is fixed to $N_s=8{,}192$.
Note that, although the conditional dynamics in Eq.~\eqref{system dynamics} is stochastic due to the mid-circuit measurement outcomes, the output in Eq.~\eqref{linear dynamics} becomes deterministic in the infinite-shot limit, since it is obtained by averaging over all possible measurement records.
This also means that, from the viewpoint of output calculation, 
the quantum dynamics of the model can be represented as a linear map acting on the ensemble-averaged system state, as follows. 
Defining the ensemble-averaged system state as
\[
\bar{\rho}_{t-1}
=
\sum_{\bm{m'}_{t-1}\in\{0,1\}^n}
p(\bm{m'}_{t-1})\,\rho_{t-1}^{(\bm{m'}_{t-1})},
\]
the output can be written compactly as
\[
\langle Z_{i,a} \rangle
=
\mathrm{Tr}\!\left[
Z_{i,a}\,
\Phi_{u_t}\!\left(
\bar{\rho}_{t-1}\otimes \sigma_a
\right)
\right].
\]
Accordingly, when the measurement outcomes are not conditioned on, 
the ensemble-averaged system state after one timestep is given by
\[
\bar{\rho}_{t}
=
\sum_{\bm{m}_t}
\mathrm{Tr}_a\!\left[
M_{\bm{m}_t}\,
\Phi_{u_t}\!\left(
\bar{\rho}_{t-1}\otimes \sigma_a
\right)
M_{\bm{m}_t}^{\dagger}
\right].
\]
Equivalently, since the measurement outcomes are averaged over, this can be written as
\[
\bar{\rho}_{t}
=
\mathrm{Tr}_a\!\left[
\Phi_{u_t}\!\left(
\bar{\rho}_{t-1}\otimes \sigma_a
\right)
\right].
\]
Before the next timestep, the ancilla is reset and prepared again in 
$\sigma_a$. 
Thus, while individual trajectories depend on the stochastic measurement outcomes, 
the ensemble-averaged evolution relevant to the output is governed by a linear 
input-dependent quantum channel acting on the system state. That is, the nonlinearity necessary in the reservoir is not intrinsic to the quantum system \cite{mujalNL, goviaNL, InnocentiNL}; rather, in our method, the nonlinearity is introduced in the encoding process \(u_t \rightarrow \hat{U}(u_t)\) .

The regression model is constructed using the matrix
\(X=[\boldsymbol{h}_{t_f}, \ldots, \boldsymbol{h}_{t_l}, \boldsymbol{1}]^T\)
which contains the reservoir output vectors at different timesteps and the all-ones vector \(\boldsymbol{1}\).
Here, \(t_f\) and \(t_l\) represent the first and last timesteps of the training phase, respectively.
Using the output weight vector \(\boldsymbol{w}_{\mathrm{out}}\), we write
\begin{align*}
    \boldsymbol{y}_{\mathrm{pred}} = X\boldsymbol{w}_{\mathrm{out}}
\end{align*}
to predict the target signal \(\boldsymbol{y}_{\mathrm{target}}\).
The optimal \(\boldsymbol{w}_{\mathrm{out}}\) that minimizes
\(\| \boldsymbol{y}_{\mathrm{target}} - \boldsymbol{y}_{\mathrm{pred}}\|\)
is obtained using the pseudoinverse \((X^T X)^{-1}X^T\), yielding
\begin{align*}
\boldsymbol{y}_{\mathrm{pred}}
    &= X(X^TX)^{-1}X^T\boldsymbol{y}_{\mathrm{target}}.
\end{align*}

We next specify the intended unitary part \(\hat{U}(u_t)\) of the channel \(\Phi_{u_t}\), which encodes the input time series.
In the 24-qubit experiments described below, we use the following product structure:
\[
    \hat{U}(u_t) = \bigotimes_{i=0}^{5}V_{4i,4i+1,4i+2,4i+3},
\]
where each \(V_{4i,4i+1,4i+2,4i+3}\) is a four-qubit unitary matrix acting on a two-qubit system with indices \((4i,4i+1)\) and a two-qubit ancilla register with indices \((4i+2,4i+3)\).
Thus, our QRC is a product of four-qubit components.
The detailed structure of \(V_{4i,4i+1,4i+2,4i+3}\) is illustrated in Fig.~\ref{fig:architecture}(d):
\begin{align*}
V_{j,k,l,m} &= \big(\mathrm{CX}_{k,m}\mathrm{CX}_{j,l}\big) U_{j,k}(u_t),  \notag\\
U_{j,k}(u_t) &= \mathrm{CX}_{j,k}
\mathrm{RZ}_k(s_{u_{t}})\mathrm{CX}_{j,k}\big(\mathrm{RX}_j(s_{u_{t}}) \otimes \mathrm{RX}_k(s_{u_{t}})\big) ,
\end{align*}
where \(s_{u_t}=au_t\) with \(a\in\mathbb{R}\) is the scaled input, and \(\mathrm{CX}_{j,k}\) denotes the CNOT gate with control qubit \(j\) and target qubit \(k\).
The gates \(\mathrm{RX}_j(s_{u_t})=\exp(-is_{u_t}X_j/2)\) and \(\mathrm{RZ}_k(s_{u_t})=\exp(-is_{u_t}Z_k/2)\) are rotation gates applied to the \(j\)-th and \(k\)-th qubits, respectively.
We employ this circuit structure to compare our scheme with the natural-noise scheme~\cite{naturalQRC}.
The circuit structures of the two schemes are essentially the same, except that the natural-noise scheme does not contain ancilla qubits for readout.
 
The present repeated-measurement scheme is closely related to the repeated-measurement QND scheme proposed in Ref.~\cite{ibm-qnd}, formulated as follows. After the input-dependent system evolution, the $n$ ancilla qubits are reset to the $\ket{0}$ state, and each ancilla qubit is then coupled to a unique system qubit, after which the ancilla qubits are measured; see Fig.~\ref{fig:architecture}(d). Let $O^{(i)}_{r,j}$ denote an observable on system qubit $i$ when $r=s$, or on ancilla qubit $i$ when $r=a$, at the $j$-th measurement. In our case, $O^{(i)}_{r,j}$ is either $Z$ or $I$. 
Let $\mathcal{W}_{u_j\cdots u_1}$ be the ordered product $\mathcal{W}_{u_j \cdots u_1} = \mathrm{CNOT}_{a,s}U(u_j)\cdots \mathrm{CNOT}_{a,s}U(u_2) \mathrm{CNOT}_{a,s}U(u_1)$, where $\mathrm{CNOT}_{a,s}$ denotes the tensor product of CNOT gates applied to each pair of system qubits and its corresponding ancilla qubit. As shown in Ref.~\cite[Appendix C]{ibm-qnd}, if 
\[
Z_a(j) = \mathcal{W}^{\dag}_{u_j \cdots u_1} (\otimes_{i=1}^{n} O^{(i)}_{a,j}) \mathcal{W}_{u_j \cdots u_1}, 
\]
then $[Z_a(j),Z_a(k)]=0$ for all $j,k$. Thus $\{Z_a(j);j=1,2,\ldots\}$ is a sequence of commuting observables in the Heisenberg picture and is therefore a sequence of QND observables. Because of this, the repeated measurement of the (multi-ancilla) observable $\otimes_{i=1}^{n} O^{(i)}_{a,j}$ at different times $j=1,2,\ldots$ yields a sequence of random variables that has a joint probability distribution and forms a discrete-time classical stochastic process (this is not the case if they are non-commuting at different times). Averaging these random variables at each measurement time labelled by $j$ over many independent runs of the repeated QND measurements gives an estimate of the quantum expectation $\langle \otimes_{i=1}^{n} O^{(i)}_{s,j} \rangle$ of the system observable $\otimes_{i=1}^{n} O^{(i)}_{s,j}$ in the Schr\"{o}dinger picture; for more details, see \hyperref[app:qnd]{Appendix~B}. We also note that the CNOT gates can be replaced with other gates between the ancilla and system qubits, as long as the ancillas are reset and prepared to the same state before the gate is applied (for example in Section \ref{subsec:measurement-strength}).  Finally, this model can be directly extended if there is decoherence such that the quantum state of the system qubits can be modeled as undergoing a Markovian evolution represented by a CPTP map $\Phi(u_k)$ acting only on the system qubits during the application of $U(u_k)$ followed by another CPTP map $\Psi_{a,s}$ when a CNOT gate is applied on the system and ancilla qubits. In this case $U(u_k)$ is replaced with $\tilde{U}(u_k)$ as a unitary dilation of $\Phi(u_k)$ and, similarly, $\Psi_{a,s}$ is replaced with its unitary dilation $\widetilde{\mathrm{CNOT}}_{a,s}$. Note that in this setting the noise acting during the application of $U(u_k)$ and the subsequent application of $\mathrm{CNOT}_{a,s}$ are independent with a factored composite quantum state.

\subsection{Benchmark task: NARMA}
\label{subsec:method_narma}

NARMA is a dynamics that takes an input time series $\{u_t\}$ and produces an output time series $\{y_t\}$.
We consider the following three types of NARMA.
\\
NARMA2:
\[
    y_{t+1} = 0.4y_{t} + 0.4y_ty_{t-1} + 0.6u_t^3 + 0.1
\]
NARMA5 and NARMA10:
\[
    y_{t+1} = \alpha y_t+\beta y_t \Biggl(\sum_{i = 0}^{d-1}y_{t-i}\Biggl) + \gamma u_{t-d+1}u_{t} + \delta,
\]
where $(\alpha, \beta, \gamma, \delta) = (0.3, 0.05, 1.5, 0.1)$ are hyperparameters and $d$ denotes the order, or memory length, of the NARMA dynamics; $d=5$ and $d=10$ for NARMA5 and
NARMA10, respectively.
In this work, we take the following input time series:
\begin{align}
u_t = 0.1\biggl(\sin{\Bigl(\frac{2\pi\bar{\alpha}t}{T}\Bigl)}\sin{\Bigl(\frac{2\pi\bar{\beta}t}{T}\Bigl)}\sin{\Bigl(\frac{2\pi\bar{\gamma}t}{T}\Bigl)} + 1\biggl), \label{eq:u_input}
\end{align}
where \((\bar{\alpha},\bar{\beta},\bar{\gamma},T) = (2.11, 3.73, 4.11, 100)\).

Our goal is to construct the QRC so that its output time series $\{\hat{y}_t\}$ well approximates $\{y_t\}$.
To evaluate the performance of the QRC for the NARMA task, we use the following normalized mean square error (NMSE):

\[
    \mathrm{NMSE_{train}} = \frac{\sum_{t=t_{ts}}^{t_{te}}(\hat{y}_t - y_t)^2}{\sum_{t=t_{ts}}^{t_{te}} y_t^2}
 \qquad
    \mathrm{NMSE_{test}} = \frac{\sum_{t=t_{ps}}^{t_{pe}}(\hat{y}_t - y_t)^2}{\sum_{t=t_{ps}}^{t_{pe}} y_t^2},
\]
where \(t_{ps}\) is the timestep at which the prediction starts and \(t_{pe}\) is the timestep at which the prediction ends, while \(t_{ts}\) is the timestep at which the training starts and  \(t_{te}\) is the timestep at which the training ends.
In Section \ref{subsec:result_narma}, we calculate \(\mathrm{NMSE_{test}}\) with \(t_{ps} = 81\) and \(t_{pe} = 100\) and \(\mathrm{NMSE_{train}}\) with \(t_{ts} = 10\) and \(t_{te} = 80\), except for NARMA10 (\(t_{ts} = 20\)). (Note that the initial 10 or 20 steps of time series data, i.e., before the timesteps \(t_{ts}\), are used for washout and are ignored.)

\subsection{Temporal information processing capacity}

To evaluate the memory and nonlinearity of our QRC system as well as the reproducibility of input processing, we employ the temporal information processing capacity (TIPC) \cite{IPC_calc,quantum_noise}, which can comprehensively characterize the information-processing mechanism performed in the QRC system.
The QRC system processes the current input $u_t$ by delaying (e.g., $u_{t-1}$) and/or nonlinearly transforming it (e.g., $u_{t}^2$).
These processes are reproducible against input series while irreproducible processing includes other elements such as time and noise and can be expressed by a product of such input term and a past state (e.g., $u_t x_{1,t-1}$).
In the framework of TIPC, the processed input terms are represented by bases $z_t^{(i)}~(i=1,2,\ldots)$ which are orthogonalized from each other to avoid overlapping the amount of processed inputs.
The capacity $C(X,z^{(i)})$ represents the amount of processed inputs $z_t^{(i)}$
via the system state $\boldsymbol{x}_t$ and is formulated as follows:
\begin{align}
    C(X, z^{(i)}) = 1 - \frac{\min_{\boldsymbol{w}}\sum_{t=1}^T(z_t^{(i)} - \boldsymbol{w}^\top\boldsymbol{x}_t)^2}{\sum_{t=1}^T(z_{t}^{(i)})^2},
\label{eq:tipc}
\end{align}
where $\boldsymbol{w} \in \mathbb{R}^N$ is the weighting vector,
$\boldsymbol{x}_t=[x_{1,t},\cdots, x_{N,t}]^\top\in\mathbb{R}^N$ is the state vector of the reservoir;
and $z_t^{(i)} \in \mathbb{R}$ is a polynomial function composed of the input history $\{u_t,u_{t-1},\ldots\}$ and past state time series $\{\boldsymbol{x}_{t-1},\boldsymbol{x}_{t-2},\ldots\}$ such as an orthogonalized term of $u_t,u_{t-1},u_tx_{1,t-1},x_{2,t}$.
The polynomials composed of only inputs (e.g., $u_t,u_{t-1}$) are reproducible for an identical input sequence, while those including the state variable (e.g., $u_tx_{1,t-1},x_{2,t}$) are irreproducible processing units (see \hyperref[app:tipc]{Appendix~A}).

The capacities were detected with a threshold calculated from a distribution of capacity error.
The statistical significance level was set at $p=5\%$, meaning that, in the case where the time-series length is $200$ ($1000$), TIPC values greater than a threshold of $0.14$ ($0.021$) were selected (see \hyperref[app:tipc]{Appendix~A}).

\begin{figure*}[t]
    \centering
    \begin{center}
     \includegraphics[width=\textwidth]{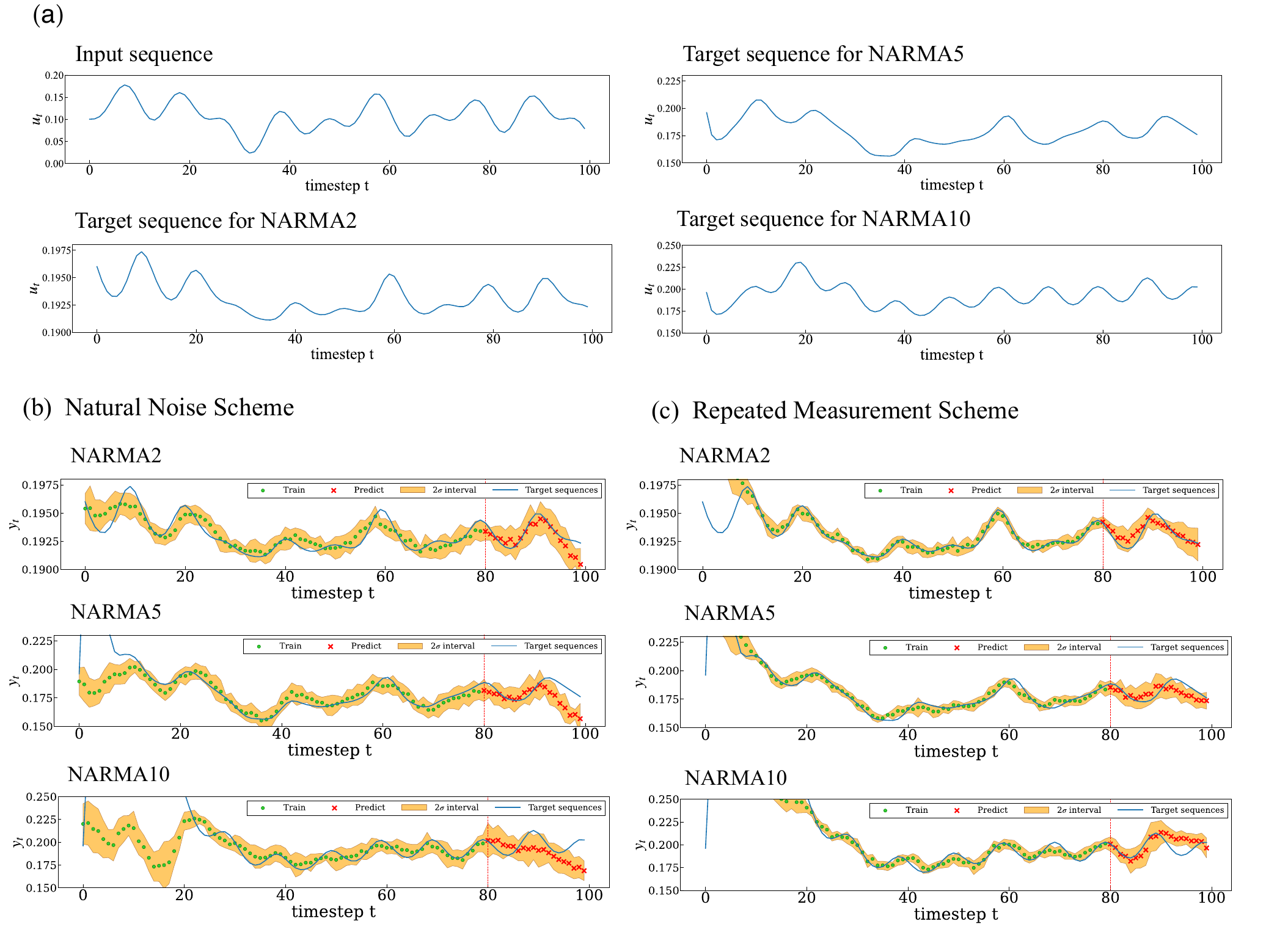}
    \end{center}
    \caption{
    Comparison of the natural-noise and repeated-measurement schemes for the NARMA tasks.
    (a) Time-series data used in the experiment: the input sequence and the target sequences for NARMA2, NARMA5, and NARMA10.
    (b,c) Learning and prediction results for the NARMA tasks obtained using (b) the natural-noise scheme and (c) the repeated-measurement scheme.
    The blue line represents the target trajectory, while the green dots and red crosses represent the averaged QRC outputs in the training and prediction phases, respectively.
    The orange shaded region represents the \(2\sigma\) range of the experimental trajectories.
    The results are averages over 10 trials for each task.
    }
    \label{fig:benchmark}
\end{figure*}

\section{Results}
\label{sec:results}
\subsection{Comparison of two QRC schemes}
\label{subsec:result_narma}

First, we compare the repeated-measurement scheme with  the natural-noise scheme \cite{naturalQRC} in terms of training and prediction accuracy as well as execution time for the NARMA benchmark tasks (NARMA2, NARMA5, and NARMA10), using ibm\_kingston.
For the natural-noise scheme, we used 12 qubits from ibm\_kingston as the system, i.e., a subset of the 156-qubit device, following the same circuit architecture as in Ref.~\cite[Results and discussion, NARMA task]{naturalQRC}.
For the repeated-measurement scheme, we also used ibm\_kingston, with 12 qubits serving as system qubits and another 12 qubits serving as ancilla qubits for measurement.
Specifically, the entire system consists of six subsystems, each implementing the four-qubit quantum circuit shown in Fig.~\ref{fig:architecture}(d).
The arrangement of the physical qubits is shown in Fig.~\ref{fig:architecture}(c), where each set of four qubits surrounded by a gray dotted line represents one subsystem.

Fig.~\ref{fig:benchmark}(a) shows the input sequence and the target output sequences used in the NARMA tasks.
The input sequence shown in the top-left panel of Fig.~\ref{fig:benchmark}(a) and defined in Eq.~\eqref{eq:u_input} is encoded into the rotation angle of the system unitary in Fig.~\ref{fig:architecture}(d), with \(s_{u_t} = 2u_t\).
Figs.~\ref{fig:benchmark}(b,c) show the time-series regression results for the natural-noise and repeated-measurement schemes, respectively, where the blue solid lines show the target trajectories.
The plotted QRC outputs are averages over 10 independently trained/evaluated trials, each using reservoir outputs estimated from 8{,}192 shots.
The green dots at timesteps 10--80 correspond to the training phase, and the red crosses at timesteps 81--100 correspond to the prediction phase.
The initial 10 timesteps are used for washout for NARMA2 and NARMA5, while the initial 20 timesteps are discarded for NARMA10, as described in Sec.~\ref{subsec:method_narma}.
The orange shaded region represents the \(2\sigma\) range of the trajectories.

Regarding execution time, the time required to obtain one output trajectory (i.e., one sample path before taking the average) was 8.1 seconds for the repeated-measurement scheme and 226 seconds for the natural-noise scheme.
These values were calculated from the actual execution times reported for jobs run on the IBM Quantum Platform, excluding queue time.
This difference suggests that our scheme has a significant advantage in terms of execution times.
This reduction arises because the repeated-measurement scheme requires only a single quantum circuit to generate one output trajectory in this experiment, whereas the natural-noise scheme requires 100 circuits.
Note that a very precise comparison of execution times is challenging because the IBM Quantum Platform system spends a significant amount of time on circuit compilation and classical processing.
With respect to regression accuracy, Figs.~\ref{fig:benchmark}(b,c) show that both schemes can fit each NARMA trajectory with small deviations over the entire time series.
In terms of prediction, the repeated-measurement scheme outperforms the natural-noise scheme, especially for NARMA5 and NARMA10.

\begin{figure}[tbp]

            \centering
              \begin{tabular}{|c|c|c|c|} \hline
                \multicolumn{2}{|c|}{Natural noise scheme} & \multicolumn{2}{c|}{Repeated measurement scheme} \\ \hline
                \multicolumn{4}{|l|}{\ NARMA2} \\ \hline
                TRAIN & TEST & TRAIN & TEST \\ \hline
                $1.01\times 10^{-5}$  & $1.48\times 10^{-5}$ & $1.93\times 10^{-6}$ & $7.10\times 10^{-6}$ \\ \hline
                \multicolumn{4}{|l|}{\ NARMA5} \\ \hline
                TRAIN & TEST & TRAIN & TEST \\ \hline
                $5.65\times 10^{-4}$ & $3.75\times 10^{-3}$ & $3.41\times 10^{-4}$ & $1.10\times 10^{-3}$\\ \hline
                \multicolumn{4}{|l|}{\ NARMA10} \\ \hline
                TRAIN & TEST & TRAIN & TEST \\ \hline
                $1.43\times 10^{-3}$ & $6.42\times 10^{-3}$ & $6.82\times 10^{-4}$ & $1.95\times 10^{-3}$\\ \hline
              \end{tabular}
              \tblcaption{NMSE values in the training and test phases for NARMA2, NARMA5, and NARMA10. NMSE values are computed from the averaged QRC output trajectories over 10 trials.}
              \label{tb:narma:table}

\end{figure}

Table~\ref{tb:narma:table} shows the NMSE values in the training and test phases, computed between the target trajectories and the averaged QRC outputs.
In the training phase, both schemes yield small NMSE values, with the repeated-measurement scheme yielding slightly smaller values than the natural-noise scheme.
This indicates that both QRCs can sufficiently fit the NARMA target sequences.
In the test phase, the repeated-measurement scheme achieves lower NMSE values than the natural-noise scheme for all NARMA tasks.
These results indicate that, overall, the repeated-measurement QRC can approximate the NARMA target dynamics in both training and prediction more accurately than the natural-noise scheme.

\begin{figure*}[p]
    \centering
    \begin{center}
    \includegraphics[height=0.83\textheight,keepaspectratio]{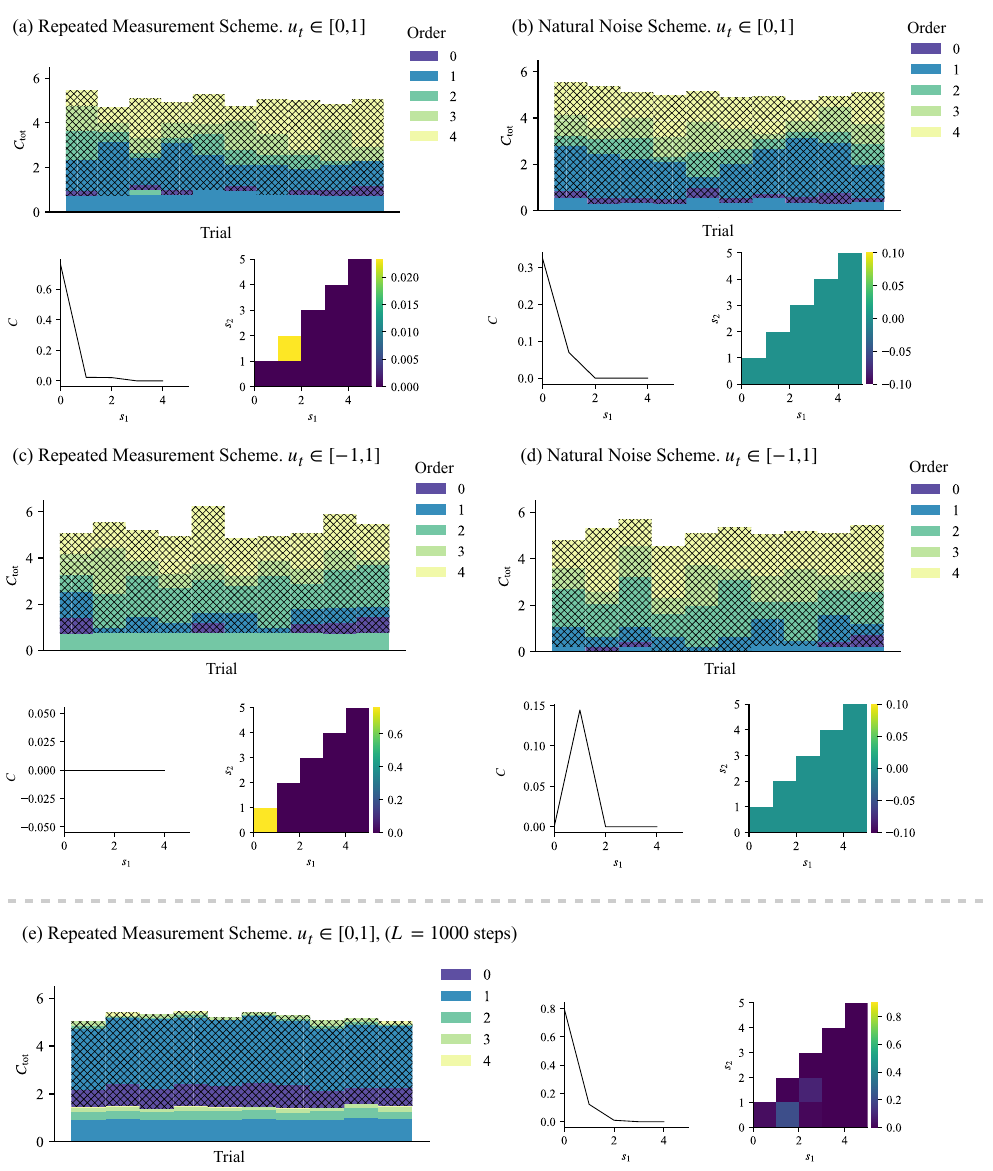}
    \end{center}
    \caption{
    TIPC values of the repeated-measurement and natural-noise schemes.
    In all subfigures (a)--(d), the upper panel shows the TIPC values from 10 trials, with colors indicating the input order from 0 to 4.
    The hatched bars represent the time-variant components of the TIPC, while the non-hatched bars represent the time-invariant components.
    The line graph in the lower-left panel shows the first-order capacity of the input, and the heatmap in the lower-right panel shows the second-order capacity; \(s_1\) and \(s_2\) represent time delays.
    (a,c) TIPC of the repeated-measurement scheme for the asymmetric-input case \(u_t\in[0,1]\) and the symmetric-input case \(u_t\in[-1,1]\), respectively.
    (b,d) TIPC of the natural-noise scheme for the asymmetric-input case \(u_t\in[0,1]\) and the symmetric-input case \(u_t\in[-1,1]\), respectively.
    All four cases are calculated using time series of length \(L=200\).
    (e) TIPC of the repeated-measurement scheme for the asymmetric-input case \(u_t\in[0,1]\), calculated using a time series of length \(L=1,000\).
    }
    \label{fig:task:continuity}
\end{figure*}

The NARMA results evaluate task-specific prediction performance. To further examine whether this improvement is associated with the intrinsic information-processing capacity of the reservoir, we next compute the TIPC in Eq.~\eqref{eq:tipc} using uniform random inputs in the symmetric case \(u_t \in [-1,1]\) and the asymmetric case \(u_t \in [0,1]\), where the input length is 200.
The results are summarized in Figs.~\ref{fig:task:continuity}(a)--(d), where in all subfigures the upper panel shows the TIPC values from 10 trials and the lower panels show the first- and second-order capacities.
The first-order capacity represents the first-order components of the input, such as \( u_t, u_{t-1}, u_{t-2} \), while the second-order capacity represents the second-order components of the input, such as \( u_t^2, u_t u_{t-1}, u_{t-1} u_{t-2} \).
The variables \(s_1\) and \(s_2\) in the lower panels represent the time delays of the input, i.e., \(u_{t-s_1}\) and \(u_{t-s_2}\).
The trial-to-trial variation in the TIPC values shown in the upper panels is small in both schemes, indicating that the experiment is reproducible.
The upper panels show that the repeated-measurement scheme has larger time-invariant TIPC components, indicated by non-hatched bars, than the natural-noise scheme for both the symmetric- and asymmetric-input cases.
That is, the repeated-measurement scheme exhibits higher TIPC-based computational capability than the natural-noise scheme.
Next, we examine the first- and second-order capacities, shown by the line graph in the lower-left panel and the heatmap in the lower-right panel, respectively.
For asymmetric inputs, the repeated-measurement scheme shows a nonzero first-order capacity. For symmetric inputs, the first-order capacity is suppressed, whereas nonzero second-order capacities appear. 
 Although the natural-noise scheme shows a first-order component in Fig.~\ref{fig:task:continuity}(b,d), its total time-invariant capacity remains smaller than that of the repeated-measurement scheme. These TIPC results support the NARMA benchmark results by showing that the repeated-measurement scheme provides larger task-independent reservoir capacities under the present experimental conditions.

We remark that, in the lower-left panel of Fig.~\ref{fig:task:continuity}(a), the time-invariant memory suddenly drops to zero as soon as a nonzero delay occurs.
However, this does not mean that the QRCs, particularly the repeated-measurement scheme, do not have memory.
This is because the TIPC in Fig.~\ref{fig:task:continuity}(a), as well as in the other subfigures of Fig.~\ref{fig:task:continuity}, is calculated using only short time series to allow comparison with the natural-noise scheme, for which the time-series length on the quantum computer was limited to 200 steps in these experiments because of the cost of circuit preparation.
As discussed in Sec.~\ref{sec:intro}, by using the repeated-measurement scheme, we can obtain longer time series and accordingly more precise TIPC estimates.
To demonstrate this, we calculated the TIPC using a time series of length 1,000 obtained via the repeated-measurement scheme.
For this long-sequence TIPC calculation, we used the CNOT-coupling setting, corresponding to the strongest coupling case discussed in Sec.~\ref{subsec:measurement-strength}.
As shown in Fig.~\ref{fig:task:continuity}(e), the TIPC includes time-invariant capacities with longer delays, indicating that the quantum reservoir has finite memory.
This is consistent with the fact that the QND measurement does not erase the information. Furthermore, we observed second-order capacity, which was not observed in Fig.~\ref{fig:task:continuity}(a).
These results suggest that using longer time series provides a more accurate characterization of the memory and nonlinear processing capacity of the repeated-measurement QRC.

\subsection{Choice of measurement strength}
\label{subsec:measurement-strength}

\begin{figure*}[t]
    \centering
    \begin{center}
    \includegraphics[width=170mm]{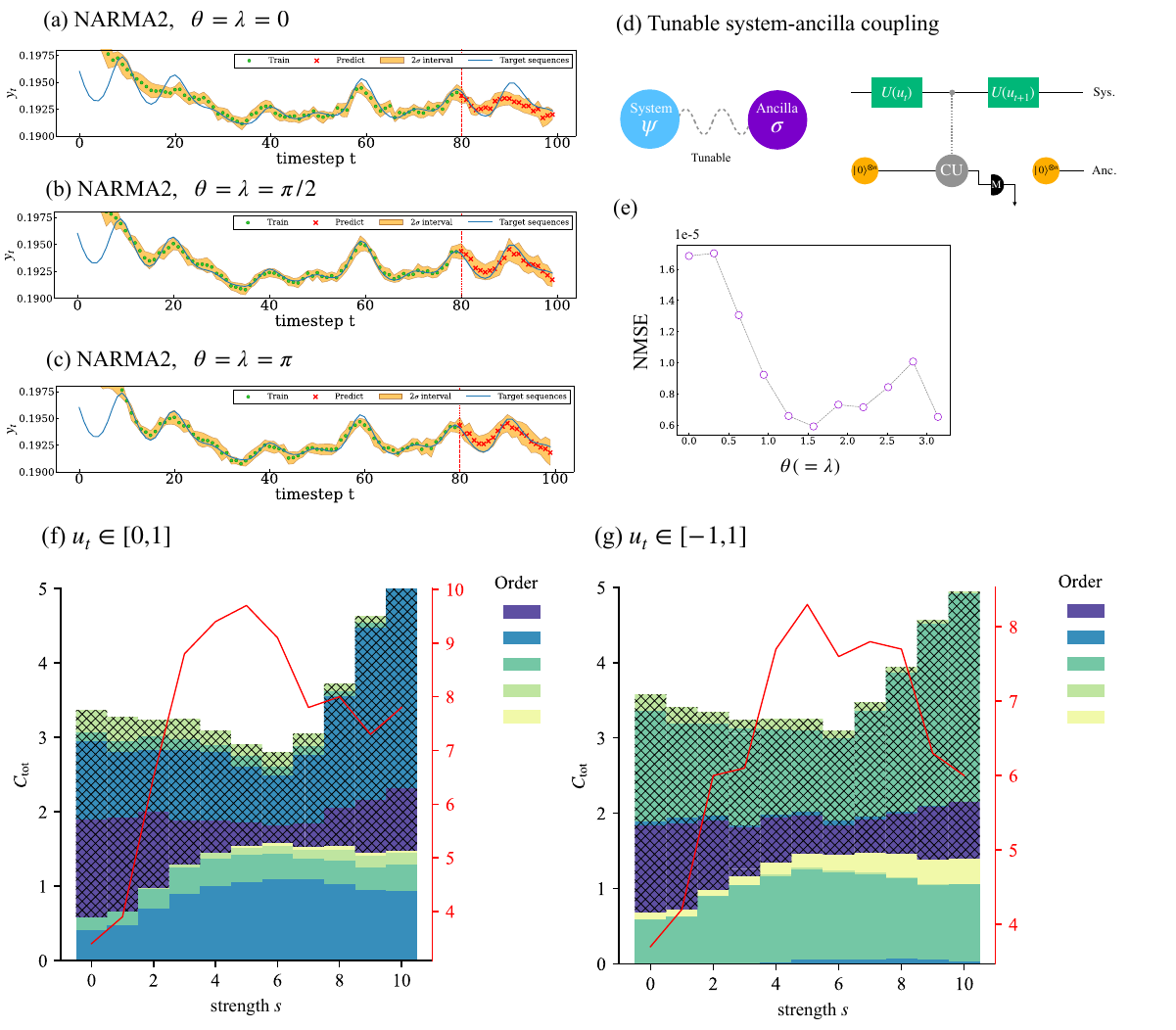}
    \end{center}
    \caption{
    Performance of the repeated-measurement scheme as a function of the interaction strength.
    (a)--(c) Learning and prediction results for the NARMA2 task with different CU-gate parameters.
    The blue line represents the target trajectory, while the green dots and red crosses represent the averaged QRC outputs in the training and prediction phases, respectively.
    The orange shaded region represents the \(2\sigma\) range of the experimental trajectories.
    (d) Schematic diagram of the tunable system--ancilla interaction, where a CU gate is used to control the amount of information extracted from the system by measurement.
    (e) Parameter dependence of the NMSE in the NARMA2 experiment. The horizontal axis represents the CU-gate parameters \(\theta\) and \(\lambda\), which are parameterized by the measurement strength \(s\) as \(\theta=\lambda=s\pi/10\).
    (f,g) TIPC values averaged over 10 trials for (f) the asymmetric-input case and (g) the symmetric-input case.
    The red line represents the richness averaged over 10 trials.
    }
    \label{fig:task:intensity}
\end{figure*}

In the repeated-measurement scheme, repeated measurements induce back-action, meaning that strong measurements may cause dissipation in the system, potentially leading to the loss of information. Thus, there is a trade-off between the amount of available information and the strength of dissipation, making it important to identify the optimal balance point where information can be extracted accurately while preserving the required memory. In fact, for the online protocol, Ref.~\cite{QRC-weak} provides an analytical bound on the optimal measurement strength in terms of the reservoir memory. In our QRC implementation, the coupling strength between the system and the ancilla can be experimentally varied, allowing us to investigate whether such a trade-off point exists. In the repeated-measurement scheme, we can replace the CNOT gates used to couple the system qubits to the ancilla qubits with controlled unitary (CU) gates as follows: 
\begin{equation} \label{eq:cnot}
\begin{split}
CU(\theta, \phi, \lambda, \gamma)
=\left(
\begin{matrix}
1 \ & \ 0 & \ 0 & \ 0 \\
0 \ & \ e^{i\gamma}\cos{(\theta/2)}  & \ 0  & \ -e^{i(\gamma + \lambda)}\sin{(\theta/2)} \\
0 \ & \ 0 & \ 1 & \ 0 \\
0 \ & \ e^{i(\gamma + \phi)}\sin{(\theta/2)} & \ 0 & \ e^{i(\gamma + \phi + \lambda)}\cos{(\theta/2)}
\end{matrix}
\right),
\end{split}
\end{equation}
where \(e^{i\gamma}\) denotes a global phase, and \(\gamma, \phi, \lambda,\) and \(\theta\) are tunable parameters.
When \(\gamma=\phi=\lambda=\theta=0\), \(CU\) reduces to the identity matrix, whereas when \(\gamma=\phi=0\) and \(\lambda=\theta=\pi\), it reduces to the CNOT gate.
In this experiment, we set \(\theta=\lambda=s\pi/10\) with \(s = 0,1,\ldots,10\) while fixing 
\(\gamma=\phi=0\) to examine how the prediction accuracy for the NARMA tasks and the TIPC value depend on the measurement strength.
As in Sec.~\ref{subsec:result_narma}, we used ibm\_kingston, a 156-qubit IBM superconducting device, with 12 system qubits and 12 ancilla qubits for measurement. 
The circuit architecture and qubit arrangement were the same as those shown in Figs.~\ref{fig:architecture} (c,d), except that the system–ancilla coupling is implemented using the CU gate defined in Eq.~\eqref{eq:cnot}.
Figs.~\ref{fig:task:intensity}(a)--(c) show the experimental results for NARMA2, indicating that the prediction accuracy improves as both \(\theta\) and \(\lambda\) approach \(\pi\).
The same trend was also observed for NARMA5 and NARMA10.
Fig.~\ref{fig:task:intensity}(e) further shows that the NMSE values generally decrease as \(\theta\) and \(\lambda\) increase.
In addition, the NMSE attains its minimum at \(\theta=\lambda=5\pi/10\).
These results indicate the existence of an optimal measurement strength for the linear-regression performance of our QRC, as quantified by the NMSE.

We further analyze the choice of measurement strength by calculating the TIPC.
Figs.~\ref{fig:task:intensity}(f,g) show the TIPC values as a function of the interaction strength \(s\), for both the asymmetric-input case \((u_t\in[0,1])\) and the symmetric-input case \((u_t\in[-1,1])\).
First, the time-invariant component of \(\Ctot\) roughly increases with the interaction strength in both input cases, but reaches a maximum at an intermediate strength.
Specifically, it peaks at \(s=6\), corresponding to \(\theta=\lambda=6\pi/10\), in the asymmetric-input case, whereas it peaks at \(s=7\), corresponding to \(\theta=\lambda=7\pi/10\), in the symmetric-input case, as shown in Figs.~\ref{fig:task:intensity}(f,g). 
Note that, in Figs.~\ref{fig:task:intensity}(a), (f), and (g), the regression still works to some extent (especially in the training phase), even when the measurement strength is 0, i.e., when the system qubits and the ancilla qubits are not explicitly coupled. We speculate that this behavior is caused by crosstalk errors between the system qubits and the ancilla qubits. Since current superconducting devices still contain substantial noise, the system and ancilla qubits can become correlated through noise between neighboring qubits even without explicit coupling.
In addition, we also calculated a quantity called richness, which quantifies the variety of input components that the reservoir can process.
Specifically, richness is defined as the number of statistically significant capacities associated with polynomials composed only of input variables.
For example, if the TIPC contains components such as \(u_t\), \(u_t^2\), \(u_t x_t\), and \(u_{t-1}x_t\), then the richness is 3, because \(u_t\) and \(u_t x_t\) are counted as the same input component, while \(u_t^2\) and \(u_{t-1}\) are counted as distinct input components.
A high richness value indicates that the reservoir output contains a wider variety of nonlinear or delayed input components, suggesting that the reservoir may be better suited to complex tasks requiring diverse capacities.
Figs.~\ref{fig:task:intensity}(f,g) show that richness follows a trend similar to that of the time-invariant component of \(\Ctot\), with both quantities peaking around \(s=5\).
Finally, regarding the time-invariant component of \(\Ctot\), the asymmetric-input case, shown in Fig.~\ref{fig:task:intensity}(f), exhibits a broad range of first- and second-order capacities, whereas the symmetric-input case, shown in Fig.~\ref{fig:task:intensity}(g), exhibits only even-order capacities.
Ref.~\cite{quantum_noise} showed that, in a QRC system with amplitude damping, asymmetric inputs generate both even- and odd-order capacities, whereas symmetric inputs generate only even-order capacities.
This parity dependence is qualitatively similar to the behavior reported in Ref.~\cite{quantum_noise}, although our device dynamics includes more general hardware noise.

These results suggest that our QRC exhibits an optimal trade-off between information extraction and dissipation.
This provides a direct way to tune the performance of QRC models through the measurement strength, which is a notable advantage of the repeated-measurement scheme over the natural-noise scheme. 
Finally, as mentioned above, Ref.~\cite{QRC-weak} provides an analytical bound on the optimal measurement strength for the online protocol in terms of the reservoir memory.
Investigating the relationship between this bound and our experimental analysis is an interesting direction for future work.

\subsection{Application to regression of soft-robot dynamics}
\label{subsec:softrobot}

\begin{figure*}[t]
    \centering
    \begin{center}
    \includegraphics[width=\textwidth]{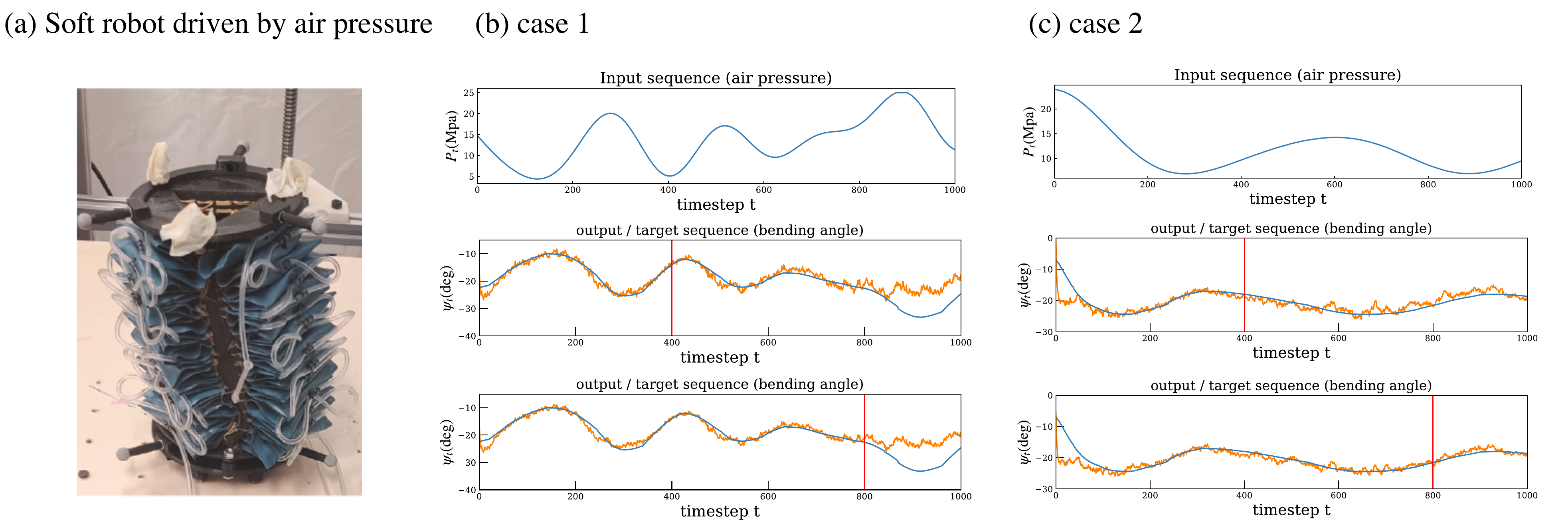}
    \end{center}
    \caption{
    (a) Soft robot used in this experiment. The robot is driven by air pressure, which is used as the input, and its bending angle is used as the target output.
    (b,c) Input sequences and target/output sequences for two different input cases. The upper panels show the input sequences, and the lower panels show the target and QRC output sequences.
    \(P_t \) and \(\psi_t\) denote air pressure and the bending angle of the soft robot at timestep \(t\) respectively.
    The blue line represents the target sequence, and the orange line represents the output of the QRC model with a trained readout.
    The middle panels show the results obtained with 400 training steps, while the bottom panels show those obtained with 800 training steps.
    In each panel, the region to the left of the red line corresponds to the training phase, whereas the region to the right corresponds to the prediction phase.
    }
    \label{fig:soft:robot}
\end{figure*}

In this section, we demonstrate the performance of our QRC model on a regression task using real-world time-series data obtained from a soft robot.
Fig.~\ref{fig:soft:robot}(a) shows the soft robot used in this experiment.
This fabric device is actuated by air pressure, causing the entire robot to bend through an air-filled pillow.
For this device, estimating the posture of the soft robot arm from the applied air pressure is a natural temporal-information-processing task, where the air pressure and bending angle serve as the input and target, respectively~\cite{soft_reservoir}.
Classical RC schemes are attractive for such tasks because of their low training cost and potential low-latency processing. Here, we examine whether the proposed QRC architecture can process this type of real-world temporal data, while also evaluating the practical runtime limitations of the present implementation.

In this demonstration, we examine whether our QRC scheme can estimate the bending angle from the nonlinear air-pressure input.
We used ibm\_kingston, a 156-qubit IBM superconducting processor, with 12 system qubits and 12 ancilla qubits.
The circuit architecture and qubit arrangement were the same as those shown in Figs.~\ref{fig:architecture}(c,d).
At each timestep, the nonlinear air-pressure input was encoded into the rotation angles of the RX and RZ gates. (In this demonstration, the rotation angles are \(s_{u_t}=0.02u_t\).)
Figs.~\ref{fig:soft:robot}(b,c) show the experimental results for two different input time series.
The blue line represents the target sequence, and the orange line represents the output of the QRC model with a trained readout.

We trained the readout using the trajectories up to timestep 400 or 800, with the initial 100 timesteps discarded as washout.
The trained readout was then used to generate predictions after timestep 400 or 800.
In each panel of Figs.~\ref{fig:soft:robot}(b,c), the region to the left of the red line corresponds to the training phase, whereas the region to the right corresponds to the prediction phase.
In all cases, our QRC appears to capture the main characteristics of the hidden dynamics to some extent.

Notably, this experiment involved \(9.83 \times 10^7\) mid-circuit measurements and the same number of reset operations \((12\ \mathrm{qubits} \times 1{,}000\ \mathrm{timesteps} \times 8{,}192\ \mathrm{shots})\), demonstrating the experimental feasibility of running the repeated-measurement circuit for a 1,000-timestep input sequence.
Compilation and execution of the circuit on ibm\_kingston, including the generation of 1,000 timesteps of data from the quantum circuit, took about 63 seconds.
Constructing the output matrix for linear regression via classical post-processing took about 5 minutes. 
Therefore, with the current implementation, our QRC scheme is not yet fast enough for practical time-series data processing and is not suitable for real-time sequential processing.
Nevertheless, these results suggest the potential of our QRC scheme for real-world applications, provided that the execution time can be substantially improved.

\clearpage
\subsection{Demonstration on devices with more than 100 qubits}

\begin{figure*}[!tbp]
    \centering
    \begin{center}
    \includegraphics[width=0.96\textwidth]{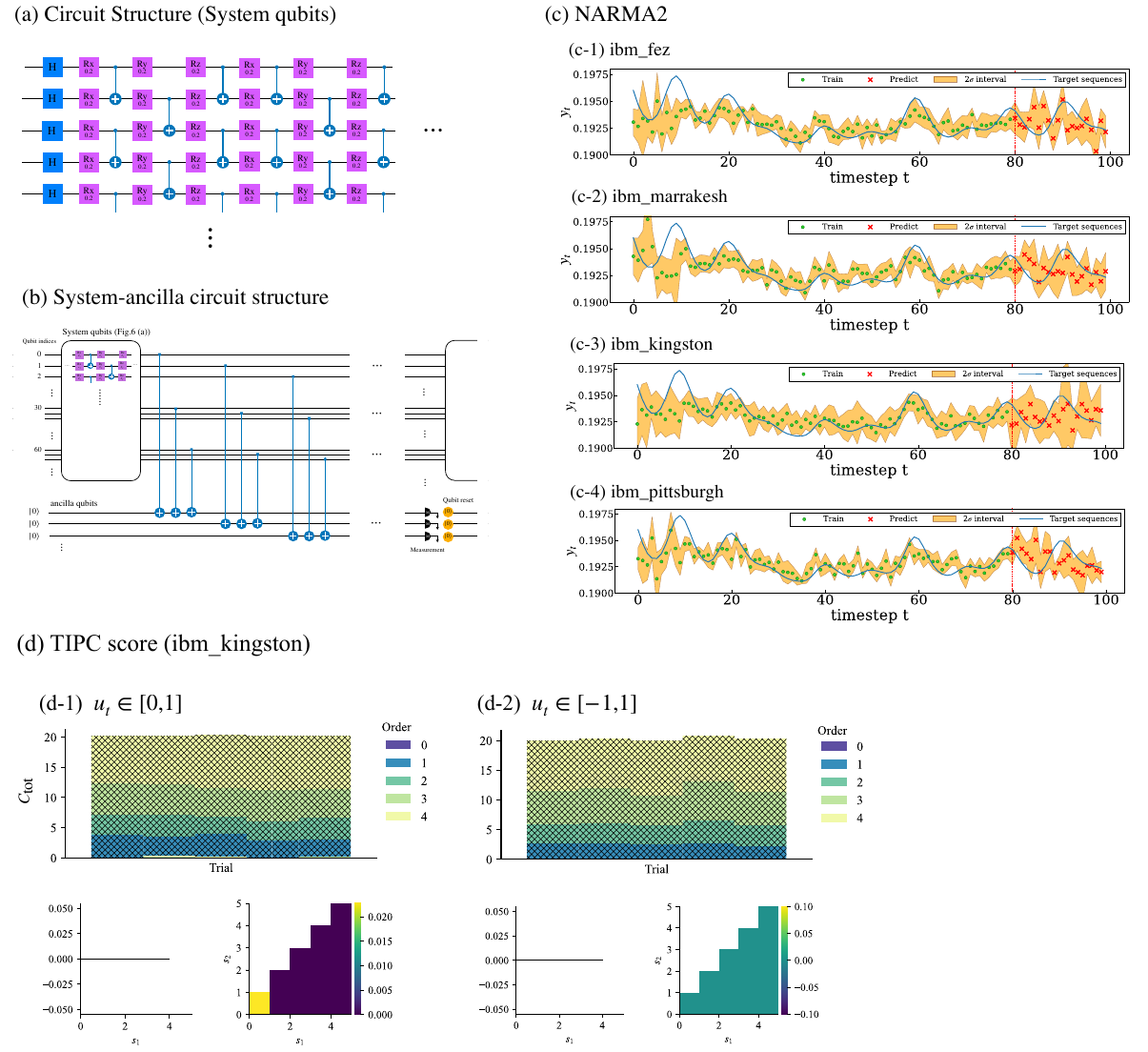}
    \end{center}
    \caption{
    (a) Circuit structure of the 90-qubit system in our 120-qubit QRC.
    (b) System--ancilla coupling circuit. Each ancilla qubit is coupled to three system qubits and then measured. As in Secs.\ref{subsec:result_narma}--\ref{subsec:softrobot}, each ancilla qubit is reinitialized by a reset operation after each measurement.
    (c) Output and target sequences for the NARMA2 task on ibm\_fez, ibm\_marrakesh, ibm\_kingston, and ibm\_pittsburgh.
    The blue line represents the target trajectory, while the green dots and red crosses represent the averaged QRC outputs in the training and prediction phases, respectively.
    The orange shaded region represents the \(2\sigma\) range of the experimental trajectories.
    The results are averages over 10 trials for each device.
    (d) TIPC values of the repeated-measurement scheme on ibm\_kingston for the asymmetric-input case \(u_t\in[0,1]\) and the symmetric-input case \(u_t\in[-1,1]\), calculated using time series of length \(L=500\).
    The horizontal axis represents the trial index.
    }
    \label{120qubit_case}
\end{figure*}

In general, QRC schemes are expected to gain computational power as their size (i.e., the number of qubits) increases.
In this section, we examine this expectation by implementing a relatively large-scale repeated-measurement QRC, composed of a 90-qubit system and a 30-qubit ancilla register, on several IBM superconducting devices with more than 100 qubits.
The circuit structure applied to the system qubits during a single timestep is illustrated in Fig.~\ref{120qubit_case}(a).
This pattern is repeated along both the qubit index and circuit-depth directions, where we applied 42 gates to each qubit along the depth direction; thus, many of the 90 qubits are expected to become entangled.
Such a large-scale circuit with substantial entangling operations may be challenging to simulate using classical computation.

Fig.~\ref{120qubit_case}(b) illustrates the system--ancilla coupling circuit between the 90-qubit system and the 30-qubit ancilla register, which differs from the structure in Fig.~\ref{fig:architecture}(d), where the system and ancilla registers have the same number of qubits.
Although several coupling patterns are possible, we adopt the structure shown in Fig.~\ref{120qubit_case}(b), in which three system qubits are coupled to each ancilla qubit for measurement.
Each ancilla qubit is reinitialized by a reset operation after each measurement, as in Fig.~\ref{fig:architecture}(d).

Fig.~\ref{120qubit_case}(c) shows the results of the NARMA task performed on four 156-qubit devices: ibm\_fez, ibm\_marrakesh, ibm\_kingston, and ibm\_pittsburgh.
In all cases, data at timesteps 10--80 were used for training, and data at timesteps 81--100 were used to evaluate prediction performance.
The initial 10 timesteps were used for washout.
As shown in Fig.~\ref{120qubit_case}(c), for all four devices, the averaged QRC outputs do not follow the target trajectory as closely as in the 24-qubit NARMA2 experiment reported in Sec.~\ref{subsec:result_narma}, in both the training and test phases. 
The NMSE values in the training and test phases for each device are summarized in Table~\ref{tab:120q_nmse}. For all four devices, both the training and test NMSE values are larger than those obtained in the 24-qubit NARMA2 experiment on ibm\_kingston.

\begin{table}[t]
\centering
\begin{tabular}{lcc}
\toprule
Device & Training NMSE & Test NMSE \\
\midrule
ibm\_fez        & $2.00\times 10^{-5}$ & $5.16\times 10^{-5}$ \\
ibm\_kingston   & $1.80\times 10^{-5}$ & $4.37\times 10^{-5}$ \\
ibm\_marrakesh  & $1.88\times 10^{-5}$ & $3.86\times 10^{-5}$ \\
ibm\_pittsburgh & $1.93\times 10^{-5}$ & $5.51\times 10^{-5}$ \\
\bottomrule
\end{tabular}
\caption{NMSE values in the training and test phases for the 120-qubit NARMA2 experiments on four IBM superconducting devices.}
\label{tab:120q_nmse}
\end{table}

To understand the cause of this performance degradation, we calculated the TIPC using data obtained from ibm\_kingston.
For each trial, we evaluated the TIPC using trajectories of length \(L=500\).
Obtaining a single trajectory required \(1.23 \times 10^8\) repeated measurements and the same number of reset operations in the 120-qubit QRC, with the total number of gate operations reaching \(10^{10}\). 
The results are shown in Fig.~\ref{120qubit_case}(d), where the horizontal axis represents the number of trials.
In contrast to the 24-qubit case shown in Fig.~\ref{fig:task:continuity}(e), no time-invariant components are observed, probably because of substantial noise.
This may partly explain the degraded performance of our QRC in both the training and test phases.
However, the presence of time-variant components in all trials indicates that the reservoir still processed the input.
As described in Sec.~\ref{sec:intro}, time-variant components in the TIPC are not directly useful for standard linear regression tasks.
However, generalized reservoir computing theory \cite{genRC} shows that time-variant components can be converted into time-invariant ones using an appropriate nonlinear readout.
Thus, an appropriate nonlinear readout within the generalized reservoir computing framework may enable us to exploit the observed time-variant components.

\begin{figure*}[!tbp]
    \centering
    \begin{center}
    \includegraphics[width=\textwidth,height=0.74\textheight,keepaspectratio]{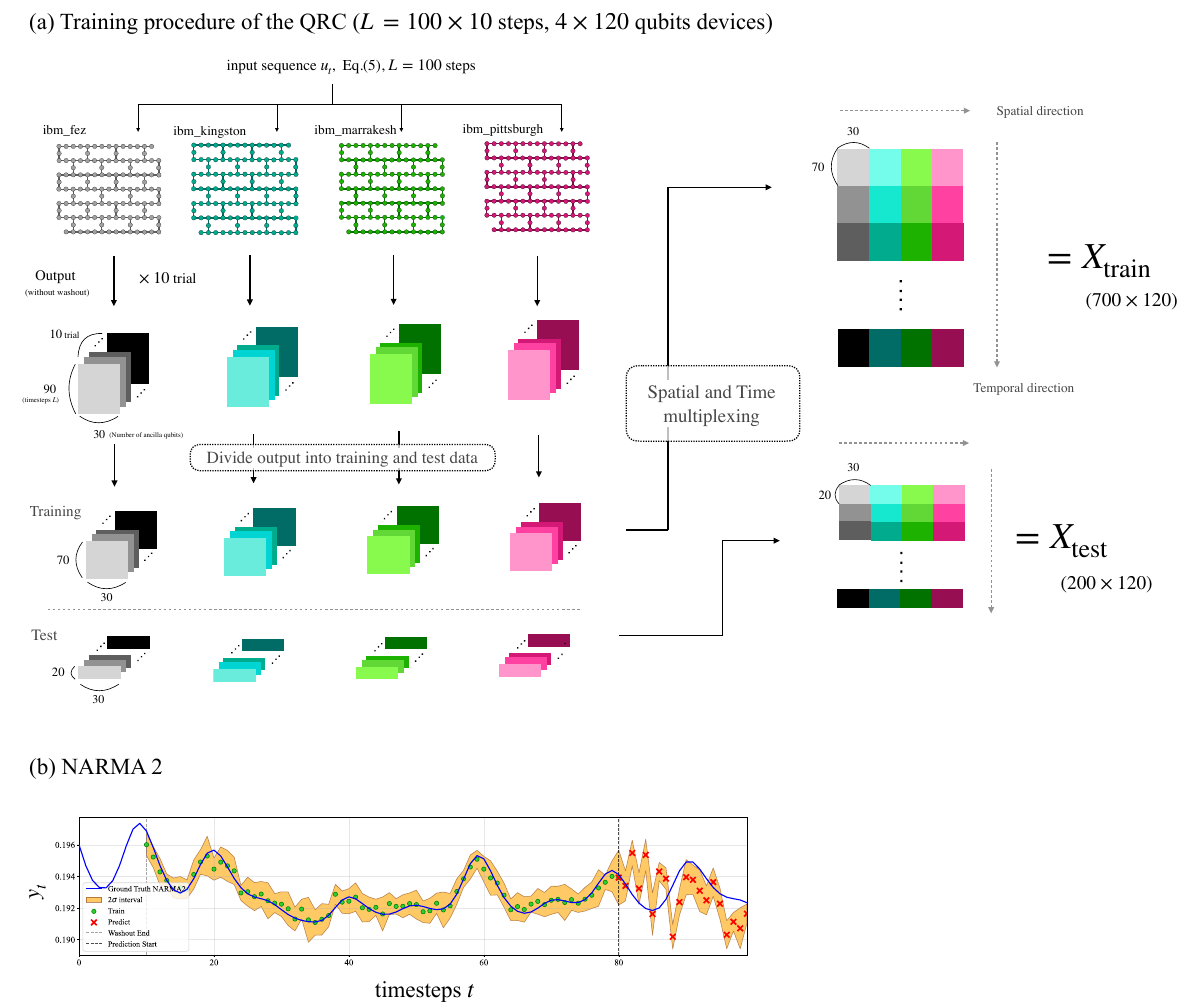}
    \end{center}
    \caption{
    (a) Training procedure for the QRC system artificially enlarged in both the spatial and temporal directions.
    Four 156-qubit devices, ibm\_fez, ibm\_kingston, ibm\_marrakesh, and ibm\_pittsburgh, are combined in both directions, corresponding to \(L=100\times 10\) timesteps and four 120-qubit QRCs.
    Each device receives the same input \(u_t\) [see Eq.~\eqref{eq:u_input}], and outputs from 10 trials are obtained from each device, as shown in the middle-left part of the figure.
    Each output is represented as a \(90\times 30\) matrix, where the rows correspond to timesteps after excluding the first 10 washout steps, and the columns correspond to the reservoir output dimension, i.e., the number of ancilla qubits.
    These outputs are then divided into training and testing sets, as shown in the bottom-left part of the figure.
    The matrix dimensions are \(70\times 30\) for training and \(20\times 30\) for testing.
    Finally, the training and testing sets are combined in both the spatial and temporal directions, resulting in output matrices \(X_{\text{train}}\) and \(X_{\text{test}}\) with dimensions \(700\times 120\) and \(200\times 120\), respectively.
    These matrices are used to train the linear readout and generate predictions according to Eq.~\eqref{eq:120q_reg}.
    (b) Output and target sequences for the NARMA2 task.
    The green dots and red crosses represent the averaged QRC outputs in the training and test phases, respectively.
    The orange shaded region represents the \(2\sigma\) range of the experimental trajectories.
    }
    \label{120qubit_4dev}
\end{figure*}

Finally, to investigate whether the computational power can be enhanced by further increasing the effective size of the QRC, we consider a procedure for artificially enlarging the system in both the spatial and temporal directions.
Specifically, to prevent overfitting caused by the increase in trainable parameters due to spatial expansion, we also expand the system in the temporal direction. Note that this procedure does not create a physically coupled larger quantum reservoir; rather, it constructs a larger classical feature matrix by concatenating outputs from independent QRC runs.
For this purpose, we used four IBM superconducting devices---ibm\_fez, ibm\_marrakesh, ibm\_kingston, and ibm\_pittsburgh---as independent QRCs and combined their outputs.
First, as illustrated in Fig.~\ref{120qubit_4dev}(a), we encoded the same input \(u_t\), defined in Eq.~\eqref{eq:u_input}, into each device and obtained the corresponding outputs.
This process was repeated independently 10 times for each device, yielding output data from 10 trials for each device.
The outputs corresponding to the first 10 input timesteps were treated as washout and were not used for training.
Next, we divided the outputs into training and testing sets, as shown in the bottom-left part of Fig.~\ref{120qubit_4dev}(a).
We then combined the divided outputs in both the spatial and temporal directions, as depicted on the right side of Fig.~\ref{120qubit_4dev}(a), to create the output matrices \(X_{\text{train}}\) and \(X_{\text{test}}\).
Thus, the outputs from 10 trials were combined in the temporal direction, while the outputs from the four IBM superconducting devices were combined in the spatial direction.
Using these two output matrices, we train the linear readout and generate predictions as follows: 
\begin{equation}
\label{eq:120q_reg}
\boldsymbol{w}_{\rm{out}} = (X_{\rm{train}}^TX_{\rm{train}})^{-1}X_{\rm{train}}^T\boldsymbol{y}_{\rm{target}}, ~~~~
\boldsymbol{y}_{\rm{pred}} = X_{\rm{test}}\boldsymbol{w}_{\rm{out}}.
\end{equation}

The target vector \(\boldsymbol{y}_{\rm{target}}\) was constructed by repeating the 70-step NARMA2 target sequence for each of the 10 trials. In Fig.~\ref{120qubit_4dev}(b), the green dots represent the averaged training trajectory, the red crosses represent the averaged testing trajectory and the orange shaded region represents the \(2\sigma\) range.
The NMSE values in the training and testing phases are \(\mathrm{NMSE}_{\mathrm{train}} = 3.18 \times 10^{-6}\) and \(\mathrm{NMSE}_{\mathrm{test}} = 7.88 \times 10^{-5}\), respectively.
Therefore, artificial enlargement of the reservoir in the spatial and temporal directions improves the training NMSE, but the test performance remains worse than that of the 24-qubit QRC on ibm\_kingston, which achieved \(\mathrm{NMSE}_{\mathrm{train}} = 1.93 \times 10^{-6}\) and \(\mathrm{NMSE}_{\mathrm{test}} = 7.10 \times 10^{-6}\), as shown in Table~\ref{tb:narma:table}. 

To conclude this subsection, our results suggest that increasing the size of a QRC does not necessarily lead to greater computational power. 
The above-mentioned negative result is consistent with recent theoretical observations on the scalability of temporal information processing with scrambling quantum reservoirs. In particular, strong scrambling can lead to concentration of measurement readouts as the reservoir size increases, while the memory of early inputs can decay exponentially with the reservoir size and reservoir iterations; in noisy settings, local noise can further induce exponential memory decay~\cite{use_disspation_2}. Although our experiment does not directly diagnose scrambling, these results suggest that simply increasing the number of qubits is not sufficient to improve QRC performance. Noise-aware architecture design, readout design, and error mitigation will be necessary for scaling repeated-measurement QRC on real devices.

\section{Relationship to other contributions}

After the earlier preprint of this work \cite{Yasuda23} was made available, several subsequent studies further developed measurement-based and open-system approaches to quantum reservoir computing and related temporal-information-processing tasks. 
Note that, in this paper, we present newly performed experiments corresponding to all the experiments reported in \cite{Yasuda23}. 
Here, we briefly summarize these developments and clarify their relation to the
present repeated-measurement QRC scheme.

\begin{enumerate}
\item
Related quantum reservoir
computing architectures based on repeated measurements and reset operations were
experimentally demonstrated on superconducting quantum processors.
In particular, Ref.~\cite{Coherence} introduced a noisy intermediate-scale
quantum reservoir computing (NISQRC) framework in which the system (memory) and ancilla qubits are coupled unitarily and the ancilla qubits are measured and reset at each timestep, enabling temporal-data processing beyond
the coherence-time limit of the hardware. We note that this is also achievable with the earlier probabilistic reset scheme of \cite{ibm-qnd} by virtue of the convergence property induced by it although the scheme has not been demonstrated experimentally due to limitations of the hardware.

At the circuit level, the architecture of Ref.~\cite{Coherence} is closely
related to the present repeated-measurement QRC scheme, in that both use
mid-circuit measurements and reset operations to extract output signals while
continuing the reservoir dynamics. The present work was originally posted
before Ref.~\cite{Coherence} and focuses on the repeated-QND-measurement
QRC architecture proposed in Ref.~\cite{ibm-qnd} but based on the natural QRC model proposed in \cite{naturalQRC} rather than the dissipative model with probabilistic reset proposed in~\cite{ibm-qnd}.  The present paper
experimentally realizes this architecture on IBM superconducting quantum
devices and investigates its computational properties.

\setcounter{enumi}{1}

\item As alluded to in the Sec.~\ref{sec:intro}, in an ideal realization of the circuit shown in Fig.~\ref{fig:architecture} in which gate operations and measurements are performed perfectly, the scheme would undergo thermalization due to the ideal scheme inducing a unital map when the ancilla qubits are traced out after they have interacted with the system qubits \cite{Coherence}. The thermalization would result in the dynamics of the circuit becoming independent of the input sequence, hence becoming useless for QRC. Despite this, our experiments indicate that this is not the case here. A plausible explanation is that the circuit is not ideal and does undergo decoherence. In this case decoherence in the actual devices is actually beneficial to preventing thermalization and enabling processing of the input sequence by the QRC. This is the same as what was observed in \cite{naturalQRC} without the use of QND measurements, where the decoherence plays a beneficial role.

\setcounter{enumi}{2}

\item
Subsequent studies proposed alternative architectures that also make essential use
of measurements, but in ways different from the repeated-measurement scheme studied
in this work.
For example, feedback-driven QRC uses projective measurements followed by
classical feed-forward operations, where the measurement outcomes are fed back into
the reservoir dynamics to help preserve or recover memory of prior inputs that would otherwise be
erased by measurement back-action~\cite{feedforward_1}.
This approach addresses a difficulty closely related to the one considered in the
present work, namely how to extract information from a quantum reservoir while
retaining memory of past inputs.
However, it does so by actively feeding the measurement outcomes back into the
system, rather than by using the repeated-QND-measurement readout adopted in
the present work.

Another related study analyzed an indirect-measurement-based online protocol in
which measurement back-action itself can enhance time-series processing when the
measurement strength is properly tuned~\cite{qnd_similar}.
More broadly, repeated  measurement-driven circuit architectures have also shown
that mid-circuit measurements and feed-forward operations can generate
nontrivial quantum dynamics even at shallow depth~\cite{feedforward_2}.
In particular, such circuits can induce nonlinearity through measurements and construct long-range entangling feature maps,
which were evaluated in a quantum reservoir-computing benchmark.
These studies highlight a broader direction in which measurements are not merely readout operations, but are used as active dynamical resources for quantum machine learning and, in some cases, temporal information processing.

\end{enumerate}

\section{Conclusion}

In this paper, we experimentally demonstrated a QRC scheme based on repeated QND measurements on IBM superconducting devices, following the theoretical proposal of Ref.~\cite{ibm-qnd}.
To evaluate the prediction accuracy and execution time of this scheme, we performed NARMA benchmark tasks and compared the results with those obtained using the natural-noise scheme.
Our QRC scheme achieved higher prediction accuracy and shorter execution time than the natural-noise scheme.
These results experimentally demonstrate that the repeated-measurement QRC is effective for regression tasks, as it avoids reinitializing the system at each timestep and reduces execution time.
Additionally, we performed experiments on a soft-robot regression task, demonstrating that our QRC can handle long time-series data with 1,000 timesteps.

We also calculated the TIPC to analyze the memory and nonlinearity underlying the computational capability of our QRC.
We showed that our QRC exhibits larger time-invariant TIPC components than the natural-noise scheme.
We also confirmed that the time-invariant component of the TIPC reaches its maximum at an intermediate measurement strength.
This corresponds to an optimal trade-off between the amount of available information and the strength of dissipation, consistent with Ref.~\cite{QRC-weak}.
Lastly, we also implemented a 120-qubit QRC, consisting of 90 system qubits and 30 ancilla qubits on IBM superconducting devices to investigate the potential computational power of large-scale QRCs that may be challenging to simulate using classical computation.
Although larger QRCs are generally expected to have greater computational power, our large-scale QRC did not outperform the smaller QRC.
The TIPC analysis suggests that this degradation may be attributed to substantial noise. This negative result highlights the importance of noise-aware architecture design rather than simply increasing the number of qubits.

Several issues remain to be examined, including how repeated measurements can be used more effectively to reduce computation time and improve performance toward practical applications.
Although our scheme achieved a speedup of several tens of times over the conventional method, further speedup should be possible.
In principle, the repeated-measurement scheme ideally has a time complexity of $\mathcal{O}(N_S L)$, compared with $\mathcal{O}(N_S L^2)$ for the conventional method. In practice, however, the observed execution time was limited by the overhead of implementing repeated measurements on current quantum devices.
Therefore, reducing the time required for repeated measurement processing is crucial for applying QRC to real-time applications.
Such investigations may help clarify the regimes in which repeated-measurement QRC offers practical advantages over conventional QRC implementations.

Analytically characterizing the measurement strength is also an important direction for future work.
In fact, Ref.~\cite{QRC-weak} discusses the existence of an optimal measurement strength in the online protocol, based on the trade-off between information extraction and preservation of quantum coherence, and provides analytical bounds on the optimal measurement strength as a function of the reservoir memory.
However, for our method, we still cannot provide analytical bounds on the optimal measurement strength under realistic hardware noise. 
Developing such analytical bounds is a promising direction for the theoretical analysis of QRC in the NISQ era. Another interesting direction is to clarify the relationship between the hardware-induced non-unital dynamics observed in the present experiments and engineered non-unital dynamics. In our implementation, the effective non-unital behavior is expected to arise from natural hardware noise and decoherence, whereas recent work has proposed controllable damping as a way to realize tunable and stable non-unital dynamics and to prevent information loss under repeated mid-circuit measurements~\cite{ctrl_damping}. Comparing these two mechanisms, or combining engineered damping with the repeated-measurement architecture studied here, may help clarify how non-unital dynamics should be designed for long-sequence QRC on quantum hardware.

The 120-qubit experiments indicate that naive scaling is strongly limited by noise. Future work should therefore combine noise-aware circuit design, optimized system–ancilla readout structures, and error-mitigation or error-suppression techniques such as those discussed in Refs.~\cite{qem-nisq, m-m3}.
Moreover, because time-variant components were observed in the TIPC analysis, investigating generalized reservoir computing \cite{genRC} with nonlinear readouts is an important future direction for exploiting such components in large-scale noisy quantum devices.

\section*{Code Availability}
The code used in this paper can be found in the \href{https://github.com/tyasuda0850/qrc_total_scripts}{GitHub repository}.

\section*{Data Availability}
The datasets used in this paper and real device information, e.g., IBM device properties at the time of the experiments and workload information, are available from the corresponding author upon reasonable request.

\section*{Acknowledgments}
This work was supported by MEXT Quantum Leap Program JPMXS0118067394, JPMXS0118067285, JPMXS0120319794, JST CREST JPMJCR2014 and JST NEXUS, Japan Grant Number JPMJNX26C5.
The results presented in this paper were obtained in part using an IBM Quantum
computing system as part of the IBM Quantum Network.
The views expressed are those of the authors and do not reflect the official
policy or position of IBM or the IBM Quantum team. The authors would like to thank Takeshi Watanabe and Toshinari Itoko at IBM Research Tokyo for their valuable technical advice and support regarding the execution of experiments on the IBM Quantum Platform. Generative AI tools were used to assist with language editing, proofreading, and drafting or revising portions of the manuscript. All scientific content, analyses, references, and the final manuscript were reviewed and verified by the authors, who take full responsibility for its contents.

\bibliographystyle{apsrev4-2-title}
\bibliography{mainRefs}

\section*{Appendix A. Temporal Information Processing Capacity}
\label{app:tipc}

To compute the temporal information processing capacity (TIPC) \cite{IPC_calc,quantum_noise}, we adopt the Volterra\text{--}Wiener\text{--}Korenberg series~\cite{polynomial_expansion} as the orthonormal polynomial expansion composed of input history and the reservoir's state history.
Let the $N$-dimensional state and input be ${\bx}_t=[x_{1,t},\cdots, x_{N,t}]^\top\in\mathbb{R}^N$ and $u_t\in\mathbb{R}$, respectively, and the state is a function of past state time series and input history, meaning that ${\bx}_t = {\bm f}(u_{t},u_{t-1},\ldots,x_{1,t-1},x_{1,t-2},\ldots,x_{N,t-1},x_{N,t-2},\ldots)$.
We consider the state matrix $X = [\bx_0,\ldots, \bx_{T-1}]^\top \in \mathbb{R}^{T\times N}$, where $r$ $(1 \leq r \leq \min \{T, N\})$ is the rank of $X^\top X$.
We obtain the normalized, linearly independent state $\hat{\bx}_t\in\mathbb{R}^r$ via the compact SVD of $X$ as ${X}={P \Sigma Q}^\top$ ($P\in\mathbb{R}^{T\times r}$, $\Sigma\in\mathbb{R}^{r\times r}$, $Q\in\mathbb{R}^{N\times r}$), where $P=[\hat{\bx}_0, \cdots, \hat{\bx}_{T-1}]^\top$.
Here, $P$ and $Q$ are real orthogonal matrices, and ${\Sigma}$ is the square diagonal matrix with non-negative real entries.
The state $\hat{\bx}_t$ is expanded as
\begin{eqnarray}
    \hat{\bx}_t &=&
    \sum_{i=1}^\infty {\bm c}_i z^{(i)}_t, \nonumber\\
    z^{(i)}_t &=& u_{t}^{n_1^{(i)}}u_{t-1}^{n_2^{(i)}}\cdots
    \hat{x}_{1,t-1}^{m_{1,1}^{(i)}}\hat{x}_{1,t-2}^{m_{1,2}^{(i)}}\cdots
    \hat{x}_{N,t-1}^{m_{N,1}^{(i)}} \hat{x}_{N,t-2}^{m_{N,2}^{(i)}}\cdots, \nonumber
\end{eqnarray}
where ${\bm c}_i\in\mathbb{R}^r$ is the coefficient vector, and
$z^{(i)}_t$ denotes the $i$-th basis in Eq.~\eqref{eq:tipc}.
Here, $N_j=\sum_{t} n_{t}^{(j)}$ and $M_j=\sum_{k=1}^r\sum_{t} m_{k,t}^{(j)}$ are the orders of inputs and reservoir internal states in this representation, respectively.

Using the Gram\text{--}Schmidt orthogonalization, we can obtain the coefficient vectors ${\bm\gamma}_i\in\mathbb{R}^r$ of orthonormalized bases $\xi^{(i)}_t$ as
\begin{align*}
    \hat{\bx}_t = \sum_{i=1}^\infty {\bm\gamma}_i \xi^{(i)}_t,~~
    C(X,z^{(i)}) = ||{\bm\gamma}_i||^2,
\end{align*}
where $z^{(i)}=[z^{(i)}_1,\cdots, z^{(i)}_{T}]^\top$, $\xi^{(i)}=[\xi_1^{(i)},\cdots,\xi_T^{(i)}]^\top$ ($||\xi^{(i)}||=1$), and
there is a one-to-one correspondence between $z^{(i)}_t$ and $\xi^{(i)}_t$.

We explain that the numerical error $C_{\rm error}(X,z^{(i)})$ of TIPC caused by the finite time length follows the $\chi^2$ distribution with $r$ degrees of freedom \cite{IPC,quantum_noise},
\begin{eqnarray}
    C_{\rm error}(X,\xi^{(i)})\sim \frac{1}{T}\chi^2(r). \nonumber
\end{eqnarray}
We choose the top $p$\% value $C_{\rm th}$ of the distribution.
We adopt $p=5$ for the quantum reservoirs implemented in real quantum machines.
Using the threshold $C_{\rm th}$, we truncated the capacity $C$ as follows:
\begin{eqnarray}
    C_{\rm truncate} =
    \begin{cases}
        C & ({\rm if}~C\ge C_{\rm th}) \\
        0 & ({\rm otherwise}) \nonumber
    \end{cases}.
\end{eqnarray}

\section*{Appendix B. Additional information on QND measurement}
\label{app:qnd}

We now give a description of the QND measurement in the Schr\"{o}dinger picture. Suppose that the system qubits are in the state $\rho_{k-1}$ just before a measurement at time index $k$. They are then coupled with the ancilla qubits by a unitary $U_{sa}$.
If the ancilla qubits are measured and give a bitstring $m_k$ as the measurement outcome (we label the outcome for each ancilla as either 0 or 1) then the unnormalised state of the system qubits after the measurement is given by the completely positive (CP) map $\mathcal{M}_{m_k}$ (equivalently, a quantum operation) defined as
$$
\mathcal{M}_{m_k}(\rho_{k-1})=\mathrm{tr}_{H_a}(P_{m_k} U_{sa} (\rho_{k-1} \otimes |0 \cdots 0\rangle \langle 0 \ldots 0|) U_{sa}^{\dag} P_{m_k}),
$$ where $H_a$ is the Hilbert space of the ancilla qubits (and $\mathrm{tr}_{H_a}$ the partial trace over $H_a$), $P_{m_k}$ is the projection operator on the ancilla qubits corresponding to the bitstring $m_k$. If between each measurement time $k-1$ and $k$ the system qubits evolve according to a unitary $U_{s,k}$ (that acts only on the system qubits, therefore as an identity operator on all ancillas) and the system qubits are initialised in the state $\rho_{0}$, while the ancillas are reset to $|0\ldots 0\rangle$ before the measurement, then at the end of the $N$-th measurement the probability of obtaining the outcomes $m_1,\ldots,m_N$ in this order is given by:
\begin{equation}
P(m_1,\ldots,m_N) = \mathrm{tr}(\mathcal{M}_{m_N} \mathcal{U}_{s,N}  \cdots \mathcal{M}_{m_{1}} \mathcal{U}_{s,1} (\rho_{0})), \label{eq:sequential-probability}
\end{equation}
where $\mathcal{U}_{s,k}$ is the superoperator $\mathcal{U}_{s,k} \rho_s =U_{s,k}  \rho_s U_{s,k}^{\dag}$ that acts only on operators of the system qubits. When the ancillas are replaced with a fresh set at every timestep $k$ the state of the system qubits post-measurement can be expressed as:
$$
\mathcal{M}_{m_k}(\rho_{k-1})=\mathrm{tr}_{H_a}(P^{(k)}_{m_k} U^{(k)}_{sa} (\rho_{k-1} \otimes |0 \cdots 0\rangle_k \langle 0 \ldots 0|) U_{sa}^{(k)\dag} P^{(k)}_{m_k}),
$$
where $P^{(k)}_{m_k}$ denotes the projector on the $k$-th set of fresh ancillas at time $k$ corresponding to the bitstring $m_k$, $U_{sa}^{(k)}$ is the version of $U_{sa}$ that couples the system qubits and the $k$-th set of ancillas, and $|0 \ldots 0\rangle_k$ is the initial state of the $k$-th ancilla qubits, each ancilla initially in the state $|0\rangle$. The probability $P(m_1,\ldots,m_N)$ is again given by Eq.~\eqref{eq:sequential-probability} but can now be re-expressed as:
\begin{eqnarray*}
\lefteqn{P(m_1,\ldots,m_N)}\\
&=& \mathrm{tr}(P^{(N)}_{m_N} U^{(N)}_{sa} U_{s,N}  \ldots  P^{(1)}_{m_1} U^{(1)}_{sa} U_{s,1} (\rho_{0} \otimes |0\ldots 0 \rangle_1 \ldots |0\ldots 0 \rangle_N \langle 0\ldots 0| \ldots _1\langle 0 \ldots 0|) \\
&&\quad U_{s,1}^{\dag} U^{(1)\dag}_{sa} P^{(1)}_{m_1} \ldots U_{s,N}^{\dag} U^{(N)\dag}_{sa} P^{(N)}_{m_N}). \end{eqnarray*}
Let
$$
j_k(P^{(k)}_{m_k})= U_{s,1}^{\dag} U^{(1)\dag}_{sa} \ldots U_{s,k}^{\dag} U^{(k)\dag}_{sa} P^{(k)}_{m_k} U^{(k)}_{sa}U_{s,k} \ldots  U^{(1)}_{sa}U_{s,1}.
$$
Observe that $j_k(P^{(k)}_{m_k})$ is the Heisenberg picture evolution of $P^{(k)}_{m_k}$ at time $k$. The probability can be written in terms of the Heisenberg picture operators as:
\begin{eqnarray*}
P(m_1,\ldots,m_N) &=& \mathrm{tr}((\rho_{0} \otimes |0\ldots 0 \rangle_1 \ldots |0\ldots 0 \rangle_N \langle 0\ldots 0| \ldots _1\langle 0 \ldots 0|) j_1(P^{(1)}_{m_1}) \ldots j_N(P^{(N)}_{m_N})).
\end{eqnarray*}
As already shown in Ref.\cite{ibm-qnd}, for this scheme $[j_k(P^{(k)}_{m_k}),j_l(P^{(l)}_{m_l})]=0$ for all $k,l$. As a consequence, $P(m_1,\ldots,m_N)$ is a proper joint probability distribution that satisfies
\begin{align}
\sum_{m_l} P(m_1,\ldots,m_l,\ldots m_N) = P(m_1,\ldots,m_{l-1},m_{l+1},\ldots, m_N) \label{eq:marginal}
\end{align}
for all $l=1,\ldots,N$, so that the sequence $m_1,\ldots,m_N$ is a realisation of a vector stochastic process (this already follows readily from the spectral theorem of functional analysis due to the commutativity of the operators $\{j_k(P^{(k)}_{m_k})\}_{k=1,\ldots,N}$; see \S 2.2 and Theorem 2.4 in the reference by Bouten \textit{et al}. \cite{QF}). That is, the measurement is QND. Note that Eq.~\eqref{eq:marginal} does not hold for non-QND sequential quantum measurements and in this case $m_1,\ldots,m_N$ is not a sample of a vector stochastic process.

Including the reset and preparation of the ancilla register into the one-step operation, we regard the evolution at timestep $j$ as an input-dependent CPTP map $\Phi_{u_j}$. Its adjoint $\Phi_{u_j}^{*}$ is defined by
\[
\operatorname{Tr}\!\left[A\,\Phi_{u_j}(\rho)\right]
=
\operatorname{Tr}\!\left[\Phi_{u_j}^{*}(A)\,\rho\right]
\]
for any density operator $\rho$ and observable $A$. 
We also define the composite channel $\Phi_{u_{j:1}} = \Phi_{u_j}\circ\Phi_{u_{j-1}}\circ\cdots\circ\Phi_{u_1}$ 
and denote its adjoint by $\Phi_{u_{j:1}}^{*}
=\Phi_{u_1}^{*}\circ\cdots\circ\Phi_{u_{j-1}}^{*}\circ\Phi_{u_j}^{*}$. 
Then the Heisenberg-picture observable associated with the $j$-th ancilla measurement is written as
\[
Z_a(j)
=
\Phi_{u_{j:1}}^{*}
\left(
\bigotimes_{i=1}^{n} O^{(i)}_{a,j}
\right).
\]
For the repeated-measurement architecture considered here, these observables satisfy $[Z_a(j),Z_a(k)] = 0$ for all $j$ and $k$. 
As a special case, in the ideal noiseless limit, the timestep evolution is described solely by the intended unitary operation together with the ancilla-system coupling. In this case, the channel $\Phi_{u_j}$ reduces to a unitary 
\[
V_{u_{j:1}}
=
\mathrm{CNOT}_{a,s}U(u_j)\cdots
\mathrm{CNOT}_{a,s}U(u_2)
\mathrm{CNOT}_{a,s}U(u_1),
\]
where $\mathrm{CNOT}_{a,s}$ denotes the tensor product of CNOT gates applied to each pair of system qubit and its corresponding ancilla qubit. 
Then, the above adjoint expression reduces to
\[
Z_a(j)
=
V_{u_{j:1}}^{\dagger}
\left(
\bigotimes_{i=1}^{n} O^{(i)}_{a,j}
\right)
V_{u_{j:1}}, 
\]
which also satisfies $[Z_a(j),Z_a(k)]=0$ for all $j,k$.
Because of this commutativity, the repeated measurement of the multi-ancilla observable $\bigotimes_{i=1}^{n}O^{(i)}_{a,j}$ at different times $j=1,2,\ldots$ yields a sequence of random variables that has a joint probability distribution and forms a discrete-time classical stochastic process. Averaging these random variables at each measurement time labelled by $j$ over many independent runs of the repeated measurements gives an estimate of the quantum expectation value $\langle \bigotimes_{i=1}^{n} O^{(i)}_{s,j}\rangle$ of the corresponding system observable $\bigotimes_{i=1}^{n} O^{(i)}_{s,j}$ in the Schr\"odinger picture. We also note that the CNOT gates can be replaced with other gates between the ancilla and system qubits, as long as the ancillas are reset and prepared to the same state before the gate is applied, as used in the measurement-strength experiment described in Sec.~\ref{subsec:measurement-strength}.

\end{document}